# Nano-Trap Engineering in MOF Microenvironment for Ultratrace Iodine Sensors

*Arun S. Babal,*[a,†] *Samraj Mollick,*[a,†] *Waqas Kamal,*[b] *Steve Elston,*[b] *Alfonso A. Castrejón-Pita,*[b] *Stephen M. Morris,*[b] *and Jin-Chong Tan*[a,*]

[a]Multifunctional Materials and Composites (MMC) Laboratory, Department of Engineering Science, University of Oxford, Parks Road, Oxford, OX1 3PJ, United Kingdom.

[b]Department of Engineering Science University of Oxford, Parks Road, Oxford, OX1 3PJ, United Kingdom.

[†]These authors contributed equally to this work.
[*]Corresponding email: jin-chong.tan@eng.ox.ac.uk

**Abstract**

Ultra-sensitive and highly selective iodine gas sensors play a crucial role during the nuclear radiation leak for a timely detection and mitigation of pollution, ensuring the safety of a vast number of operators and subsequent integrity of the facility. Herein, we rationally designed a metal-organic framework (MOF) that exhibits an outstanding performance with an almost billion-fold enhancement in the electrical response due to its optimized hydrophobicity, which allows the easy migration of iodine molecules though the channels and the presence of suitable interaction sites, temporarily anchoring the target molecule for ultra-trace sensing. The prototype sensor tested in demanding environments demonstrates its high selectivity, ultra-trace parts per billion (ppb)-level sensitivity, good reversibility, and a very fast response time even at high frequencies compared to existing adsorbents, including commercially available materials. Further, the iodine sensing at the atomic level was studied in detail by measuring the electrical response of a single crystal and, the optimal thickness of the MOF layer was identified for an industrially-viable prototype sensor by using inkjet printing. In a wider perspective, we propose a general strategy towards electrically efficient sensing materials with hybrid functionalities for engineering high-sensitivity iodine sensors for a safe and sustainable future.

**Introduction**

Nuclear is the backbone of greenhouse gas-free energy, providing up to 10% of the global energy demand.[1] To meet the ever-increasing energy demand of society, it is considered as the most promising alternative source over fossil fuels due to its high energy density, low operational cost, and low greenhouse gas emission, which leads to a reduced carbon footprint.[2,3] Despite these advantages, in developed countries, the age of nuclear energy starts to fade away due to the concerns associated with the emission of radioactive gases during nuclear accidents and fuel reprocessing that has severe long-term impacts on both the environment and human health.[4] Mostly, radionuclides are emitted in the form of gases and can enter our food chain through contaminated air, which deposits these radioactive molecules on agricultural soil and into drinking water supplies. One of the major gases of concern is iodine, with its isotopes including $^{131}$I (half-life: 8 days) and $^{129}$I (half-life: 1.7 million years), which adversely impact human metabolism and are a major cause of thyroid cancer.[5-8] Conventional iodine sensors have certain shortcomings that must be tackled, i.e. low sensitivity, poor selectivity, short-term reusability, and sensing only at higher temperatures.[9-12] To address these long-standing challenges, a more effective technology needs to be developed to selectively detect the ultra-trace levels of radionuclides in the case of a nuclear breach or in fuel preprocessing facilities so as to maintain a hazard-free working environment and to ensure safety for the wider community.

Metal-organic frameworks (MOFs) are ideal candidates for sensing studies, which have started to gain the attention of the scientific community due to their high surface area, tunable pore size, design versatility as well as with their exceptional combination of structural, thermal, and chemical robustness. MOF materials can be tailor-made to incorporate specific interaction sites to selectively accommodate target gas molecules.[13-16] Currently, optical techniques are widely studied in the field of gas-phase sensing in MOFs, but for industrial applications, a direct electric response far outweighs it due to the real-time highly compact setup with low-cost manufacturing, high readability, and better assimilation with modern electronics.[11,17-19] However, electrical sensing of non-polar gases such as $I_2$ remains an immense challenge due to the shortfall of effective materials to afford practical designs. In this regard, a material with a judicious combination of optimal hydrophobicity and interaction sites could be an ideal candidate for ultra-trace (ppb level) $I_2$ detection in the gas phase, but this has yet to be achieved.

Herein, a series of prototype sensors was prepared by drop-casting, in-situ growth of single crystals, or inkjet printing MOFs, on a platform of interdigitated electrodes (IDEs) with the aim of fabricating a highly selective and ultra-trace (ppb-level sensitivity) gas-phase iodine sensor. A ppb-level high sensitivity was realized in a MOF material through a rational tuning of the optimal hydrophobicity and guest-host interaction sites (Scheme 1). In the presence of iodine, a very fast reversible million-fold enhancement in alternating current (AC) signals and a remarkable billion-fold enhancement in direct current (DC) was achieved for the electrical response of the prototype sensor due to its high adsorption capacity. Here, the optimal hydrophobicity allows adsorption of the non-polar iodine gas molecules and suitable interaction sites selectively trap them to the host framework, accomplishing a ppb-level sensitivity. Furthermore, the gas phase ultra-high selectivity of the MOF material was tested by studying the cyclic electrical response in different volatile organic compounds (VOCs) and water-saturated environments, in addition to the optimal thickness of the MOF layers on the prototype sensor to standardize the system for potential industrial application. Finally, a room temperature electrical response study was conducted to simulate a real-time sensing application.

**Results and Discussion**

The prototype sensor platform consists of interdigitated comb-like electrode (IDE) structures on a glass substrate coated with an insulated polymeric film. For the qualitative comparison between the sensors, the electrical response (impedance, capacitance, and phase angle) from the empty IDE was measured in addition to their response toward the iodine gas, which remained unchanged as shown in Figure 1. Later, thin films of different MOFs were prepared on the sensors *via* various methods i.e., drop casting, in-situ single crystal growth, and inkjet printing (controlled layered deposition) and then dried in the presence of inert $N_2$ gas (see SI section 1.2). Powder X-ray diffraction (PXRD) was carried out for all the MOFs to confirm their crystalline structure, before characterizing the electrical response from the prototype sensors (see Figure S2). The variance in the electrical response of empty and MOF-deposited sensors was non-existent, which confirms the insulating nature of MOFs (as a low-*k* dielectric). The effect of hydrophilicity, hydrophobicity, and interaction sites of candidate MOF materials on high selectivity and ultra-trace sensitivity toward the specific gas was monitored to study their individual impact, and subsequently those individual parameters are combined to ascertain the final optimized material.

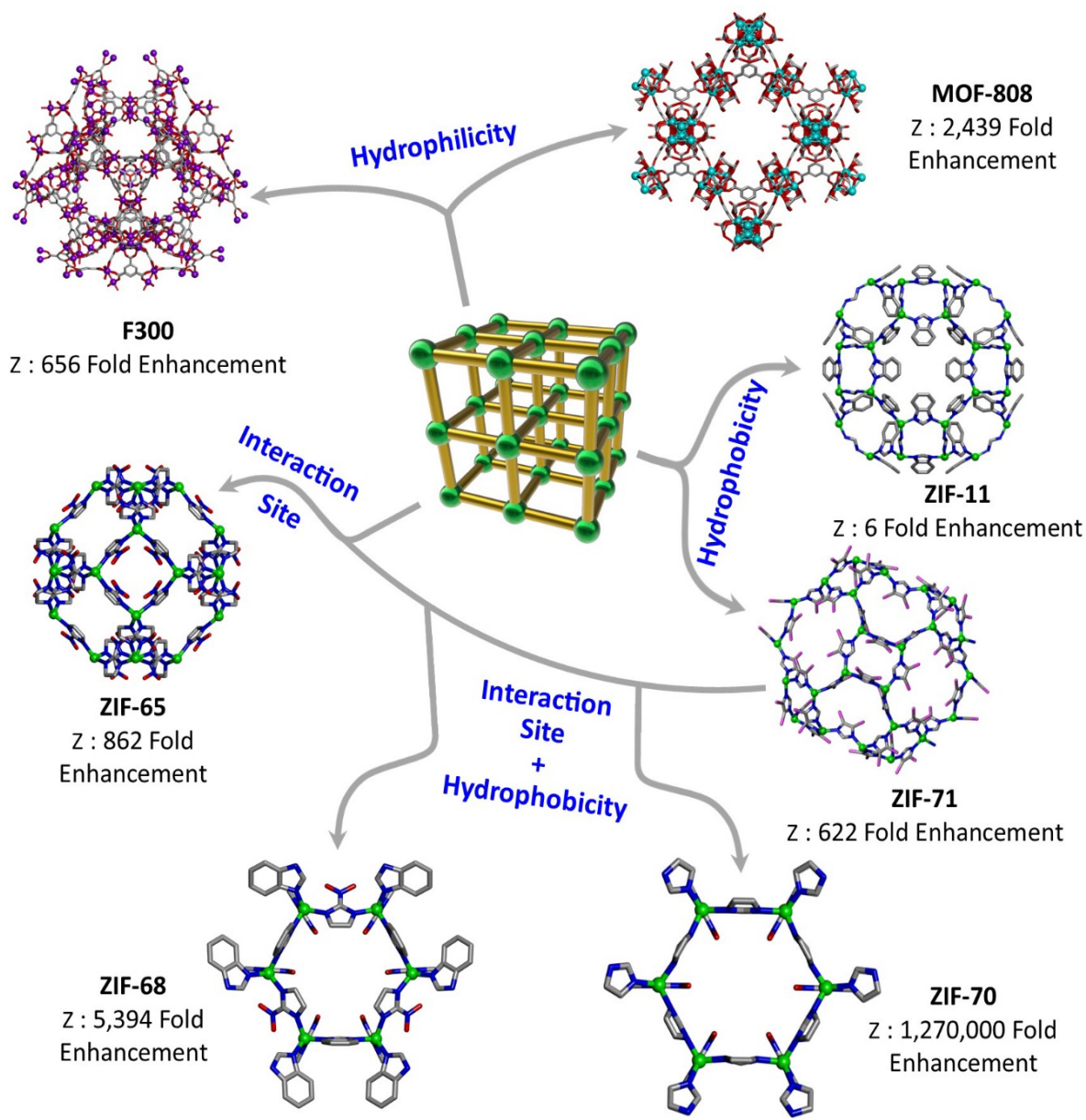

**Scheme 1: Summary of optimal chemical and physical properties to yield highly selective and sensitive MOF for gas phase iodine sensing. The *Z* number denotes the enhancement in the electrical response achieved by each of the MOF structure relative to the "empty" IDE substrate at 4 Hz (i.e. MOF-free IDE as a control sensor).**

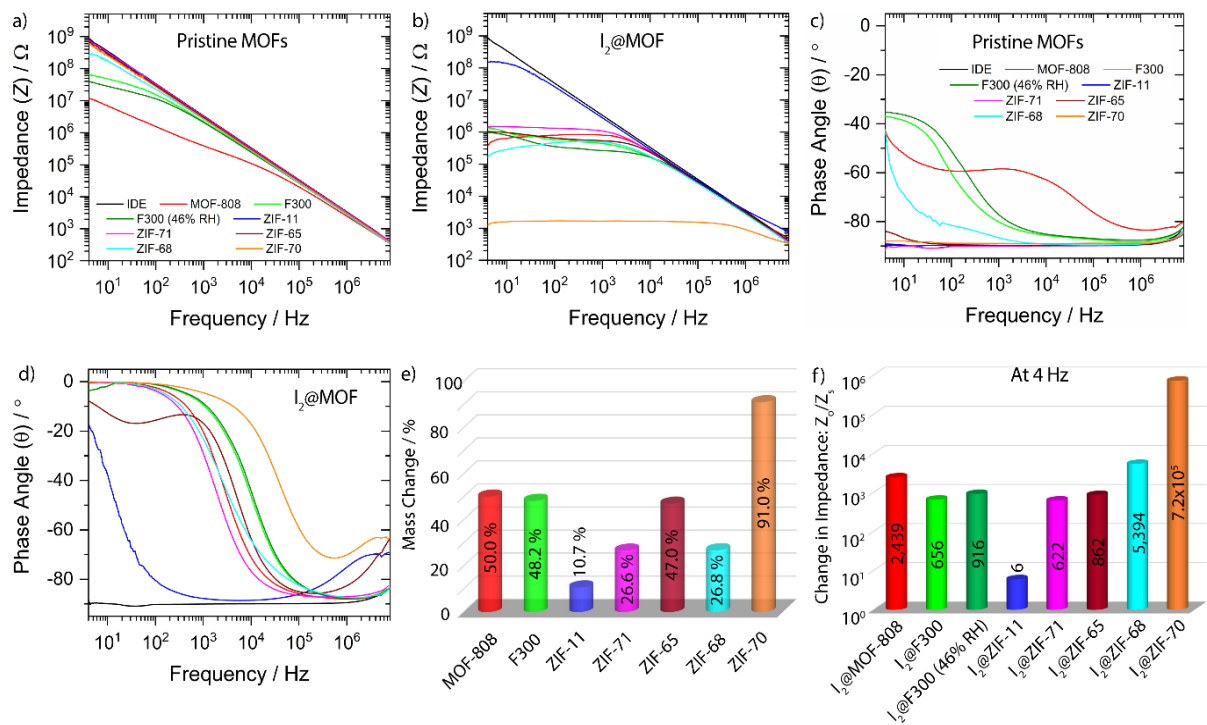

**Figure 1:** Electrical response from the MOF@IDE prototypes at room temperature: before and after iodine adsorption: (a), (b) Impedance signal and (c), (d) phase angle as a function of frequency, respectively. (e) Percentage change in sample mass, and (f) the change in impedance ratio of various MOF structures, after iodine exposure.

The different prototypical MOFs: MOF-808 and F300 for hydrophilicity, ZIF-11 and ZIF-71 for hydrophobicity, ZIF-65 for its functional nitro-group, and ZIF-68 and ZIF-70 for their interaction site and optimal hydrophobicity were selected as a sensing material for gas phase detection of $I_2$ (see Figure 1 (a)-(d) and Figures S4-S7). Unlike the other MOFs, which showed a lack of $H_2O$-induced contribution in the sensor response, the hydrophilic MOFs: MOF-808 and F300 exhibited a significant shift in electrical response at room conditions in contrast to the empty IDE sensor, suggesting that the high conductivity and polarizability of the $H_2O$ molecules absorbed on the hydrophilic sites of the porous framework greatly influence the sensor performance. To evaluate the impact of moisture on the sensor response, F300@IDE was tested at 2 different relative humidity conditions (35 and 46% RH), validating that higher electrical response is associated with the increasing RH (see Figure 1 (a) & (c) and Figure S4).

In the presence of iodine gas, the overall sensitivity of hydrophilic prototype MOFs: MOF-808 (mass change of ~50%) and F300 (mass change of ~48%) is a combination of both $H_2O$ and $I_2$ molecules (present in the MOF pores), decreasing the impedance response further by 656 and 2,439 folds, respectively (see Figure 1 (f)) and showed a noticeable transition in the phase angle (θ) from -42° (MOF-808) and -37° (F300) to almost 0° at lower frequencies, indicating a transition in electrical response from a capacitive to resistive type (see Figure 1 (d)). The slightly higher sensitivity of F300 MOF over MOF-808 can be associated to its higher iodine adsorption capacity (see Figure 1 (e)). The presence of $H_2O$ molecules in the hydrophilic pore negatively affects the inbound target molecules (shrinks the adsorption capacity of framework) and decreases the overall sensitivity as well as reusability of the sensor devices, rendering them ineffective for the real time gas phase sensing except for application as a humidity sensor. In consideration of these factors and to eliminate the impact of moisture on sensor sensitivity, we selected ZIF-11 and ZIF-71 as a prototype hydrophobic MOF for $I_2$ sensing, which showed no alterations in electrical response under ambient conditions (see Figure S5). Despite the advantages of not being affected by the moisture, the hydrophobicity causes a decrease in the iodine adsorption, which in turn leads to an unwanted lowering of the electrical response of the sensor. Unlike the extreme hydrophobicity of the ZIF-11, which brings about a decline in the iodine adsorption (mass change of ~11%) and, causing an exceedingly low enhancement in the impedance ratio of only 6-fold; the ZIF-71 framework has a comparatively higher iodine adsorption capacity (mass change of ~27%), enhancing its electrical response to 622-fold, which is comparable with the hydrophilic MOF sensors (see Figure 1 (e) & (f)).

In addition to the characterization of the hydrophilicity and hydrophobicity of the MOF candidates, the interaction sites can also play a key role in selective capture of the targeted molecules. In this regard, ZIF-65 was chosen to evaluate the impact of interaction sites on iodine adsorption and its electrical sensing response, because it has a nitro-group as part of its linker molecule. The ZIF-65@IDE is unaffected by the presence of moisture, and it adsorbs a notable amount of iodine molecule yielding a mass change of 47%, when exposed to the iodine gas (see Figure 1 (e)). The electron deficient nitro-group acts as an anchor point for the inbound iodine molecules and partially polarizes them. An increase in the polarization causes a significant rise in the capacitance value (1,876% enhancement) as seen in Figures S8 and S9, which is comparable to the hydrophilic MOFs, where water is highly polarizable. For ZIF-65 MOF, the phase angle showed a slight deviation in the pattern compared to the other tested MOFs, by not showing a complete transition from almost -90° to 0°, when exposed to iodine gas. We reasoned that the effect resulted from the interaction of iodine molecules with the framework, producing a marked 862-fold enhancement in the ratio of impedance values at 4 Hz, which is higher than the values of the hydrophobic MOF ZIF-71 as well as the hydrophilic MOF F300 (see Figure 1 (f)). Collectively, the data obtained from the hydrophilic, hydrophobic, and interaction site-based MOFs indicate that to achieve an ultra-high sensitivity $I_2$ sensor, the individual properties of MOFs are insignificant. An interaction site-based MOF with optimal hydrophobicity, which is a combination of extreme hydrophobicity and hydrophilicity could be a way forward to harness the MOF potential in the field of electrical sensing for iodine gas detection.

For this purpose, two different prototypical MOFs: ZIF-68 and ZIF-70 were chosen, comprising the same interaction site (nitro-group from 2-nitroimidazole) and different hydrophobicity based on the secondary linker molecule. The hydrophobicity of MOFs was adjusted by varying the secondary linker molecule, from the imidazole (Im) to benzimidazole (bIm) for ZIF-70 and ZIF-68, respectively. The presence of the additional benzene ring in the structure makes the ZIF-68 framework more hydrophobic compared with the ZIF-70 structure, and can provide further validation to the optimal hydrophobicity requirement theory for the ultra-high electrical gas sensing of non-polar gases. The hydrophobic linker molecule provides the protection from the highly polar water molecules by impeding their negative impact on sensor performance, while the electron deficient nitro-group acts as a temporary anchor point for the inbound iodine molecules, by partially polarizing them causing an overall shift in the structural properties to generate a very high electrical response from the prototype sensors. The iodine adsorption capacity of ZIF-68 MOF (mass change ~27%) was relatively low compared to the ZIF-70 MOF (mass change ~91%), which was a direct result of its higher hydrophobic nature (see Figure 1 (e)). In the presence of iodine gas, the increase in impedance ratio of 5,394-fold for ZIF-68 is impressive and indicative of the formation of a percolation network when exposed to a small amount of iodine, albeit this is nowhere near the astonishing increase of 0.73-million-fold in impedance ratio at 4 Hz for ZIF-70, which is an exemplar of an optimal combination of hydrophobicity and interaction site (see Figure 1 (f)). We proposed that the adsorbed iodine provides new and faster charge transfer pathways in the framework causing a decrease in the material impedance. It is well established that at higher iodine concentration upon adsorption in framework pores, they form an interconnected networks of polyiodides, which decrease the material resistance by providing the charge transfer pathways.[20,21] Likewise, for the other measurement for ZIF-70 MOF, such as the phase angle transition from -90° to 0° was stable up to < $10^3$ Hz, implying its potential as a frequency-dependent sensor material for iodine gas. From the systematic studies above, it was established that the ZIF-70 MOF, which contains both interaction site and optimal hydrophobicity is the unrivalled candidate, and it should be studied in detail to further optimize the different sample preparation parameters and operational conditions that could impact its gas phase iodine sensing performance.

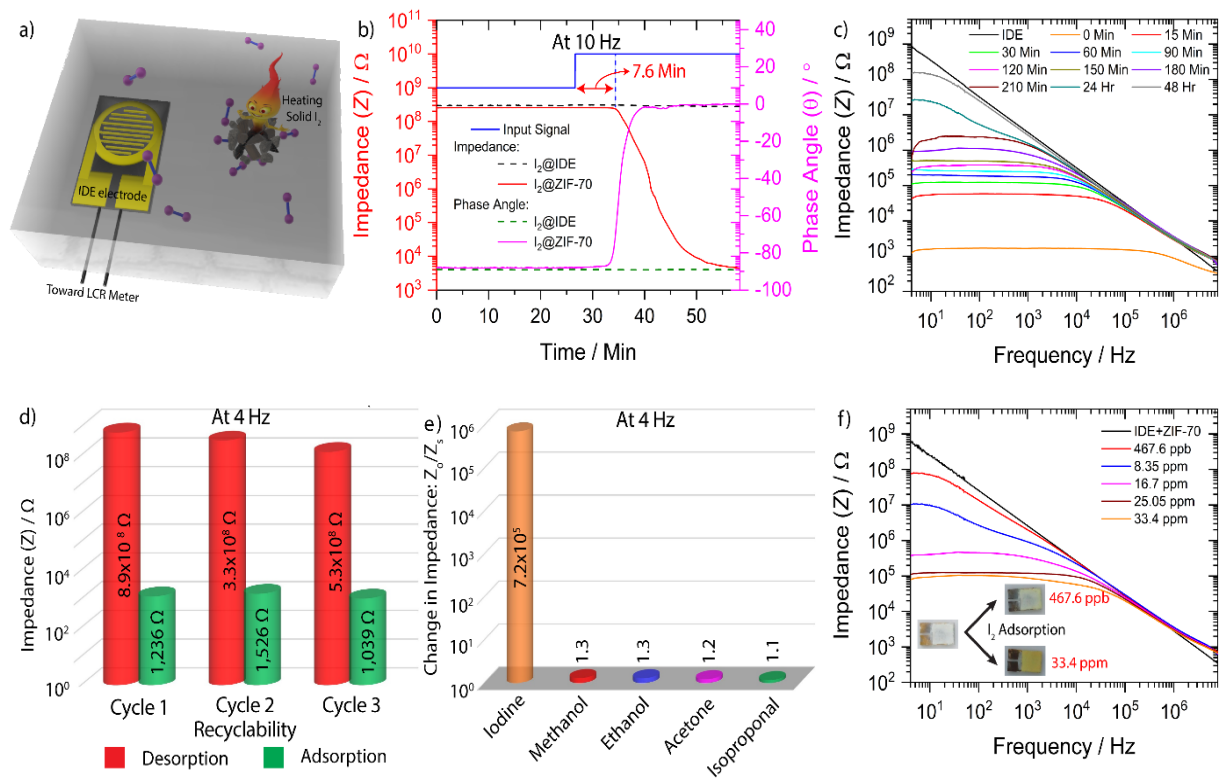

Figure 2: Drop-casting of ZIF-70 on an IDE sensor platform for iodine gas sensing: (a) A scheme of real-time iodine sensing setup. (b) Continuous impedance and phase angle measurements as a function of time, during iodine exposure, and (c) the effect of time-dependent iodine desorption on impedance value of ZIF-70@IDE as a function of frequency. (d) Recyclability study of prototype sensor over several adsorption-desorption cycles. (e) The change in impedance ratio of various MOF structures, after iodine exposure. (f) The impedance sensitivity of ZIF-70 prototype sensor at different concentrations of ppb-ppm levels.

To confirm whether the location of iodine is inside or outside the framework pore, we conducted a controlled experiment, where the guest is designed to be encapsulated inside the MOF pore. In this context, the ZIF-71 MOF was chosen due to the ease of encapsulation of the triethylamine (TEA) guest molecule inside the pore. From Figure S10, it is evident that the impedance value of ZIF-71 is much lower compared to that of the guest-encapsulated framework and, it keeps decreasing with increased guest concentration suggesting that the iodine inside the pore plays a major role in reducing the impedance value. So, it can be stated that the electrical response obtained for ZIF-70 is contributed by iodine present both on the MOF surface and inside the pores.

Advancing forward, to reveal the true performance of ZIF-70 MOF, we devised a real-time sensing environment by exposing the prototype sensor to the iodine gas at room temperature and continuously collecting the sensor transient electrical response: impedance and phase angle at 10 Hz frequency as shown in the scheme of Figure 2 (a). The electrical response of an empty IDE sensor (without MOF coating) to iodine was also recorded for comparison, which remained unaffected by the presence of iodine gas. After iodine exposure, it takes 7.6 min to register an electrical response from the prototype sensor, which can be observed in Figure 2 (b). In addition to the iodine adsorption, a time-dependent iodine desorption response from ZIF-70@IDE was also measured at room temperature. It is evident from the Figure 2 (c) that, the adsorbed iodine starts to desorb from the framework structure as soon as it is removed from the iodine-rich environment. For the initial first few hours, the desorption rate was much higher, but this decreased as a function of time. After 48 h, most of the iodine had desorbed, causing a significant drop in impedance value, which is closer to the impedance value of iodine-free MOFs, suggesting a disruption in percolation pathway. To further validate the sensor reversibility, the impedance measurement was carried out for ZIF-70 at 4 Hz for several iodine adsorption (at 70 °C for 30 min)-desorption (at 70 °C for overnight) cycles, which remained quite stable and shows a highly reversible electrical response (see Figure 2 (d)). It further strengthens the case of adopting ZIF-70 as a sensing material. Another important factor that can severely impact the sensor performance is its selectivity toward the targeted molecule. For this purpose, we investigated hydrophilic MOF F300, hydrophobic MOF ZIF-71, ZIF-65 MOF with interaction site and ZIF-70 with optimal hydrophobic-interaction site, to determine their solvent-dependent sensitivity for practical use.

As shown in the Figure S12, an enclosed chamber was specifically designed to carry out the vapor saturation study. The prototype sensors were first flushed with nitrogen gas to remove moisture content, subsequently the nitrogen was flowed through the solvent containing bubbler to form a saturated environment to carry out the electrical measurements. It was clear from Figures S13 and S14 that, except for ZIF-70 (see Figure 2 (e)), all other MOFs show an enhanced capacitance and impedance at lower frequencies in the presence of common solvent vapors, which will limit the iodine gas sensing under real-world conditions. Furthermore, sensitivity of a sensor is also a key criterion to evaluate its performance, which is dependent upon the adsorbent gas concentration. In this regard, we performed impedance measurements on ZIF-70 MOF at different iodine concentrations to validate its potential for sensing the ultra-trace amount of $I_2$ in gas phase, as shown in Figure 2(f). It was observed that even at a ppb-level concentration the material shows a clear transition from a white appearance to a yellow color, and the impedance change of the ZIF-70 MOF is still significantly higher, which confirms its capability to accomplish ppb-level sensing of iodine gas and further strengthens the case for deploying ZIF-70 as an ideal iodine gas sensing material combining ultra-high sensitivity and excellent selectivity. Additionally, an adsorption-desorption cyclic study of solvent vapor was carried out on ZIF-70@IDE at 10 Hz frequency to confirm the repeatability of the electrical response (see Figure S15).

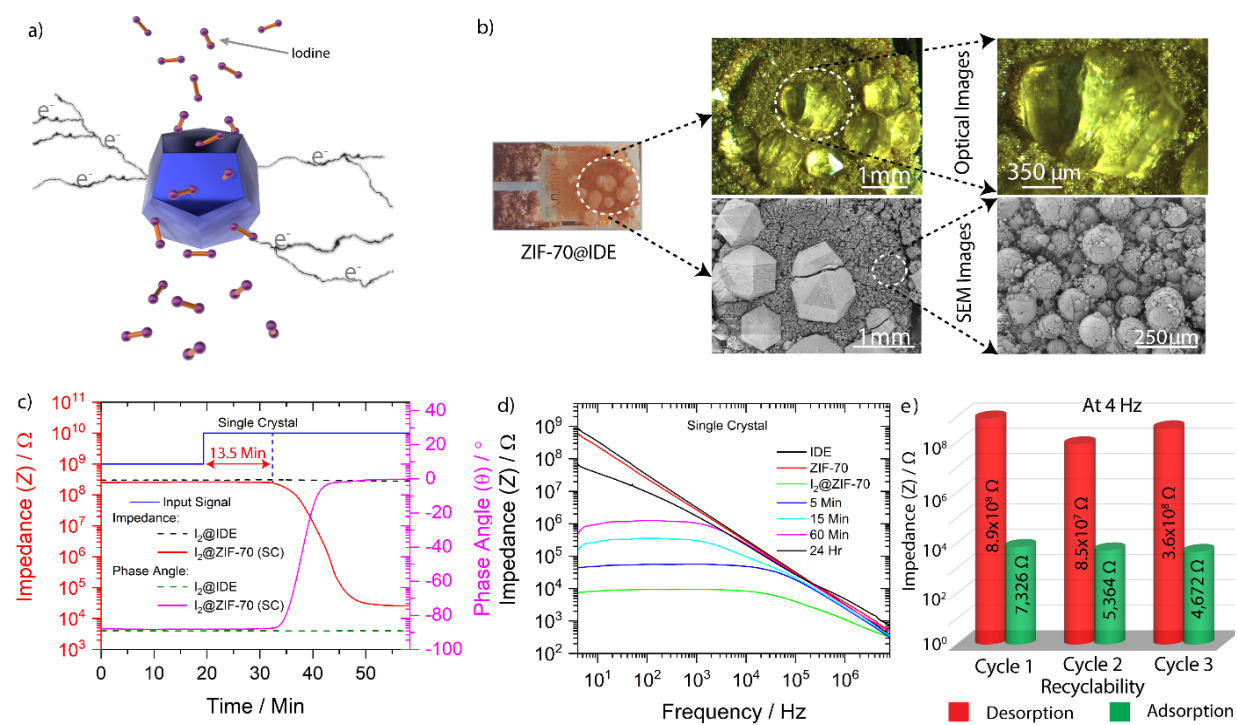

**Figure 3: Single crystal-based (SC) iodine gas sensing of ZIF-70@IDE: (a) A scheme of increase in electron conduction in single crystal resulted from iodine adsorption. (b) Optical and SEM images of ZIF-70 crystals grown on an IDE sensor platform, observed at different magnifications. (c) A real-time measurement of the transient signals of impedance and phase angle at 10 Hz frequency for ZIF-70 crystals, when exposed to iodine gas. (d) A time-dependent desorption study of ZIF-70 crystals. (e) A cyclic adsorption and desorption data to demonstrate the sensor reusability.**

For a comprehensive understanding of $I_2$ gas phase sensing in ZIF-70 framework structure, we carried out the single-crystal sensing experiments by directly growing large single crystals of ZIF-70 (> 200 μm) onto the top surface of IDE sensor chips, as shown in Figure 3 (b). Interestingly, as compared to the drop-casted sensors, the electrical response of the prototype single-crystal (SC) sensor is lower due to the crystal thickness, which delayed the electrical response by extending the $I_2$ adsorption time. It was further validated by measuring the time-dependent electrical response of the crystals in the presence of $I_2$ gas, which showed a time lag in registering the response of about 13.5 min compared to the 7 min observed for the drop-casted sensor (see Figure 3 (c)). The single crystal showed a negligible electrical response under room conditions, which can be attributed to residual solvents trapped in the pores. In the presence of iodine vapor, the single-crystal sensor showed a remarkable increase of 0.12-million-fold in impedance ratio at 4 Hz. It was also observed that the $I_2$ desorption rate of single crystal is much higher than the drop-casted sensors and can be associated with stress-induced cracks on the crystal surface. It was further supported by the reversibility experiment, in which several adsorption-desorption cycles were carried out and it showed a systemic decline in the impedance of the $I_2$ adsorbed sensor, suggesting a crack-induced enhancement in iodine adsorption. To the best of our knowledge, we have demonstrated for the first time gas phase sensing using MOF single crystals integrated on an IDE sensor platform with an unprecedented sensitivity.

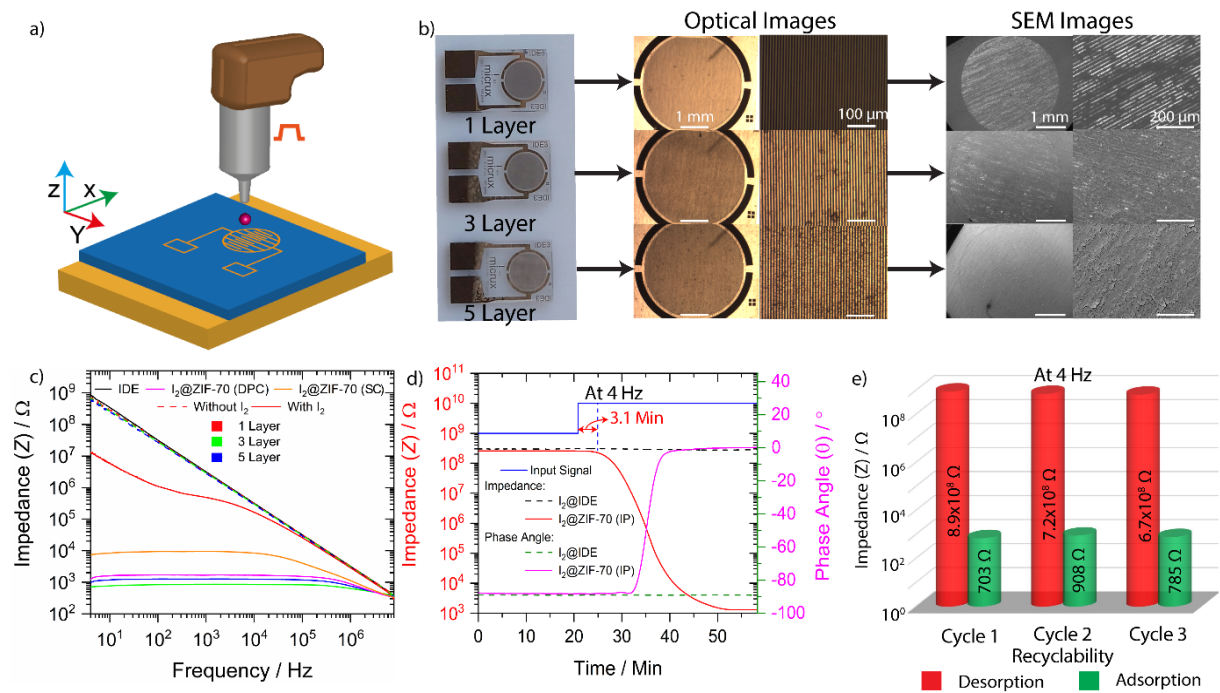

**Figure 4:** Inkjet printing (IP) of ZIF-70 on an IDE sensor platform for iodine gas sensing: (a) An illustration of the inkjet printing technique. (b) Optical and SEM images of inkjet-printed prototype IDE sensors. (c) Frequency-dependent change in impedance values for inkjet-printed sensors, before and after iodine adsorption, where DPC stands for drop-casted technique and SC for single-crystal method. For 3-layer-ZIF-70@IDE: (d) A real-time transient measurement of impedance and phase angle at 10 Hz frequency, and (e) cyclic adsorption and desorption data for testing sensor reusability for 4 Hz AC.

Inkjet printing offers several benefits over the commonly used printing techniques, which include superior print resolution, higher control of drop volume, ability to deposit material in multiple layers, scalable, mask less, and additive in nature. To push the study forward towards real-world sensor implementation, we standardized the thickness of the deposited MOF thin film by using piezoelectric-driven drop-on-demand inkjet printing (IP). In this regard, three different prototype sensors were prepared by varying the number of deposited layers (1, 3, and 5 layers) of ZIF-70 as described in the SI (section 1.2). As can be seen from Figure 4 (b) and Figure S17 that, unlike the 1-layer of ZIF-70, the 3-layer sample has completely and uniformly covered the IDE sensor surface with a film thickness of *ca*. 8 μm. In the presence of the $I_2$ gas, the electrical response from the 1-layer ZIF-70@IDE is insignificant, whereas the 3-layer ZIF-70@IDE showed a record-breaking enhancement of 1.27 million-fold due to its full coverage of the IDE and the minimum film thickness, which allowed the $I_2$ to adsorb more rapidly and efficiently. This was further supported by two experimental observations: (i) The impedance values of 5-layer ZIF-70@IDE showed a decline, which can be associated with the increased film thickness as compared to the 3-layer ZIF-70@IDE, causing an increase in the adsorption time (see Figure (c)), (ii) A time-dependent electrical response measurements of 3-layer ZIF-70@IDE, which showed a detection time of 3.1 min, revealing its rapid and efficient $I_2$ adsorption (see Figure (d)). The 3-layer ZIF-70@IDE was also examined for its reversibility through adsorption-desorption test cycles, revealing the remarkably stable electric response of the inkjet-printed films (see Figure (e)). Furthermore, the inkjet-printed MOF film showed an astonishingly high 2.84-billion-fold enhancement in the electrical response in direct current (DC), which is considerably higher than the drop-casted and single-crystal MOF@IDE responses discussed above (see Figure 5 (b)). To the best of our knowledge, this value is the highest reported enhancement in the iodine detection sensitivity of both the DC and AC frequencies (see Figure 5(c) and (d)), signifying a new record in the field of $I_2$ adsorbent sensors.[22-25]

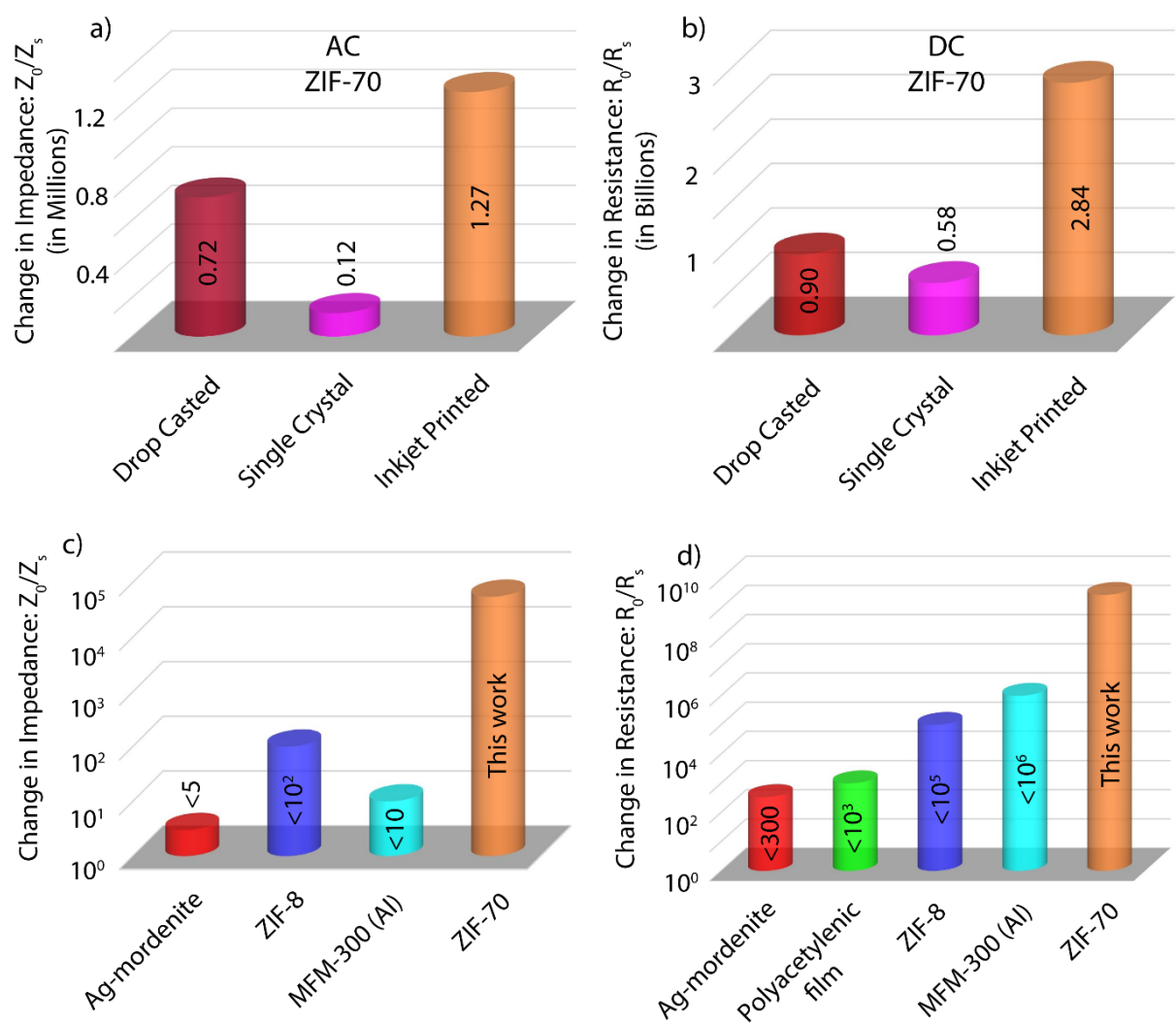

**Figure 5:** A comparative bar plot for the impedance-dependent sensitivity response of drop-casted, single-crystal, and inkjet-printed sensors in (a) 4 Hz alternating current (AC), and (b) direct current (DC). A comparison of the sensor sensitivity between this work and the previously reported IDE-based iodine sensors (data summarized in Table S5)[22-25] for sensing performance in (c) AC at 50 Hz to emulate Great Britain's national grid frequency and (d) at DC.

The post-capture phase was analyzed by PXRD, Fourier Transform Infrared Spectroscopy (FTIR), Raman spectroscopy and solid-state UV-visible diffuse reflectance spectroscopy (DRS) (Figures S19 and S20). The structural integrity remained intact after several sensing experiments confirmed by the unaltered PXRD patterns for both the powder sample and ZIF-70 single crystals, matching the simulated XRD pattern (see Figure S19 (a) and (b)). In the presence of iodine, a shift in the characteristic vibrational bands of ZIF-70 MOF was also observed in the FTIR spectra (see Figures S19 (c) and S20), indicating its ultra-high sensitivity towards the detection of iodine molecules. Unlike the blue shift in the infrared (IR) mode observed at 1174 cm$^{-1}$ (Figure S20 (h)), most of the characteristic peaks exhibit a red shift in peak positions along with peak broadening and a noticeable increase in the intensity of some vibrational bands, which are in line with literature reports.[26]

The FTIR spectra was also measured for a time- and temperature-dependent iodine desorbed ZIF-70 MOF to show the reversibility of the peak shift. The results suggest that the iodine confined in the framework is interacting with its surrounding, forming a percolation network thus causing a decline in the overall impedance. The Raman spectra was also recorded to confirm the existence of the iodine species. Two distinct vibrations were observed for the iodine adsorbed ZIF-70 MOF, which are not present in its pristine state (see Figure S19 (d)). The Raman peak at 112 cm$^{-1}$ was associated with the existence of polyiodide, whereas another peak at 166 cm$^{-1}$ was ascribed to the weakly coordinated iodine.[27,28] The underlying higher area of 166 cm$^{-1}$ peak over its counterpart suggests that iodine exists in the solid form rather than in an ionized form. For bandgap determination, the absorbance spectra for ZIF-70 MOF were measured by DRS and transformed *via* the Kubelka–Munk (KM) method to estimate the optical band gaps. We established the presence of iodine in the pore channels causes a significant reduction in the framework bandgap from 2.78 eV (pristine ZIF-70) to 1.35 eV (I$_2$@ZIF-70), which can be seen in the inset of Figure S19 (e), further substantiating the observed decrease in the impedance of the prototype MOF@IDEs described above.

**Concluding remarks**

To accomplish a highly selective and ultra-trace sensitive electrical sensing of iodine gas, we provided a mechanism to identify the key combination of physico-chemical properties in a nanoporous material for the task. The outstanding performance of the prototype MOF@IDE sensors demonstrated here challenges the existing technologies, it brings about a paradigm shift in the field of iodine sensing (see Table 1). We show a reversibly selective and even an ultra-trace ppb-level I$_2$ gas sensor, effective both at lower (Hz) and higher frequencies (MHz), making it suitable for easy integration into commercial AC electronics. This breakthrough was achieved by optimizing the level of hydrophilicity-hydrophobicity in the MOF material by leveraging the guest-host interaction sites. The mechanism exploits the tunable hydrophilicity-hydrophobicity level in a nanoporous framework, for instance ZIF-70, facilitating an uninterrupted migration of the inbound target gas species through the pores, and the presence of interaction sites as a temporary anchor point to confine gas molecules to allow chemosensing.

In the wider perspective, our work opens the door for the design of smart electrical gas sensors employing optimal hydrophobicity and interaction sites for real-time selective sensing of hazardous and toxic chemicals at ultra-trace levels. This sensing methodology can safeguard against accidents and chemical leaks in numerous industrial and consumer settings, meeting the ever-increasing demand for engineering advanced thin-film sensing technologies with supersensitive target molecule detection.

**Acknowledgements**

A.S.B. thanks the Engineering Science Studentship (EPSRC DTP – Samsung) for supporting this DPhil project. J.C.T. and S.M. thank the ERC Consolidator Grant PROMOFS (grant agreement 771575) for funding the research. J.C.T. acknowledges the EPSRC IAA award (EP/R511742/1) for additional support. A.A.C-P was supported by The Royal Society through a University Research Fellowship (URF\R\180016) and the John Fell Fund, Oxford University Press, *via* a Pump-Priming grant (0005176). We thank the Research Complex at Harwell (RCaH) for the provision of TGA, UV-Vis and Raman spectrometers.for additional support. A.A.C-P was supported by The Royal Society through a University Research Fellowship (URF\R\180016) and the John Fell Fund, Oxford University Press, *via* a Pump-Priming grant (0005176). We thank the Research Complex at Harwell (RCaH) for the provision of TGA, UV-Vis and Raman spectrometers.

## Author contributions

A.S.B., S.M. and J.C.T. designed the research. A.S.B. and S.M. conducted the materials synthesis and characterization, and A.S.B. performed the impedance measurements and data analysis, under the supervision of J.C.T. W.K. performed the inkjet printing experiments, under the supervision of A.A.C-P, S.E. and S.M.M. A.S.B. and S.M. wrote the original draft of the manuscript with input from J.C.T. All the authors contributed to the final version of the manuscript.

## Additional information

Supplementary data to this article are available.

## Competing financial interests

The authors declare that they have no known competing financial interests.

# *Supporting Information*

## *for*

# Nano-Trap Engineering in MOF Microenvironment for Ultratrace Iodine Sensors

*Arun S. Babal,[a,†] Samraj Mollick,[a,†] Waqas Kamal,[b] Steve Elston,[b] Alfonso A. Castrejón-Pita,[b] Stephen M. Morris,[b] and Jin-Chong Tan[a,*]*

[a]Multifunctional Materials and Composites (MMC) Laboratory, Department of Engineering Science, University of Oxford, Parks Road, Oxford, OX1 3PJ, United Kingdom

[b]Department of Engineering Science University of Oxford, Parks Road, Oxford, OX1 3PJ, United Kingdom

[†]These authors contributed equally to this work.
[*]Corresponding email: jin-chong.tan@eng.ox.ac.uk

## Contents

# 1. Materials preparation and characterization methods

## 1.1 Synthesis of MOF materials

Apart from F300 (purchased from Sigma Aldrich), the rest of the MOF materials employed in this study (ZIF-11, ZIF-65, ZIF-68, ZIF-70, ZIF-11, MOF-808) were synthesized by following the previously reported methods found in literature, with slight modifications.[1-5] All reagents and solvents were commercially available and used as received, sourced from Fisher Scientific, Alfa Aesar, and Fluorochem, depending on their availability.

**ZIF-11:** 2 mmol of benzimidazole was dissolved in 10 mL of ethanol and 8.5 mL of toluene, followed by the addition of ammonia hydroxide (2 mmol $NH_3$) under stirring at room temperature. After that, 1 mmol of zinc acetate dehydrate was added and stirred for the next 3 h at room temperature. The product of ZIF-11 was collected by centrifugation and washed with 50 mL of ethanol and dried at room temperature in the open air overnight.

**ZIF-65:** 0.5 mmol zinc acetate was soluble in 5 mL of DMF and rapidly mixed into 1 mmol of 2-nitroimidazole in 5 mL methanol under vigorous stirring. After 24 h of stirring at room temperature, the sample was centrifuged at 10,000 rpm to collect the product. The product was further washed three times with a copious amount of *N*, *N*-dimethylformamide (DMF) and methanol to remove the excess reactants and dried at 90 °C overnight.

**ZIF-68:** 0.5 mmol of 2-nitroimidazole, 0.16 mmol of benzimidazole and 0.5 mmol of $Zn(NO_3)_2 \cdot 6H_2O$ were mixed in 2 mL of different DMF solution separately. After that, three different solutions were combined and heated in a capped vial at 130 °C for 96 h and left to cool for 12 h. The mother liquor was decanted and the products were washed with excess DMF for four times.

**ZIF-70:** 0.36 mmol of 2-nitroimidazole, 0.36 mmol of imidazole and 0.36 mmol of $Zn(NO_3)_2 \cdot 6H_2O$ were mixed in 2 mL of different DMF solution separately. After that, three different solutions were combined and heated in a capped vial at 130 °C for 96 h and left to cool for 12 h. The mother liquor was decanted and the products were washed with excess DMF for four times.

**ZIF-71:** 2 mmol zinc acetate was soluble in 50 mL of methanol and rapidly mixed into a 50 mL methanol solution of 8 mmol of 4,5-dichloroimidazole under stirring. The mixed solution transformed from clean to turbid after a few seconds. After 24 hours of stirring at room temperature, the sample was centrifuged at 10,000 rpm to collect the product. The product was further washed three times with a copious amount of methanol to remove the excess reactants.

**MOF-808:** Trimesic acid (210 mg, 1 mmol) and zirconyl chloride octahydrate (970 mg, 3 mmol) were dissolved in DMF/formic acid (30 mL/30 mL) and placed in a large screw-capped glass jar, which was heated to 130 °C for two days. A white precipitate of MOF-808 was collected by filtration and washed four times with 400 mL of fresh DMF. The DMF-washed compound was then immersed in 100 mL of acetone for four days and during this time the acetone was replaced two times per day to facilitate the solvent exchange process. The acetone-exchanged sample was then evacuated at room temperature for 24 h and at 150 °C for 24 h to yield an activated sample.

## 1.2 IDE sensor preparation

The prefabricated thin-film gold IDEs on a glass substrate were purchased from Micrux (ED-IDE3-Au), each sensor chip contains 184 pairs of gold microelectrode with a width and a gap size of 5 μm, respectively. Before sample deposition, the IDE electrodes were rinsed with isopropanol (HPLC grade, Sigma-Aldrich, ≥99.9%) and then dried under the nitrogen gas. The impedance value of empty IDEs was measured beforehand to perform a qualitative comparison between the sensitivity of different MOF samples. In the case of ZIF-70 MOF, for drop-casting and inkjet printing, the synthesized crystals were first broken into fine particles using a tip sonicator and then deposited on the IDE electrode.

**Drop-cast method:** As described in supplementary §1.1, the MOF materials were synthesized and left in the solvent suspension to avoid aggregation. The MOF suspension was pipetted out and drop casted on the active area of the IDE in such a way that on average ~1 mg of material was deposited for each of the MOF samples. Afterwards, the IDE was dried under ambient conditions. The amount of the deposited MOF was monitored using a high-resolution balance.

**Single crystal:** To prepare a single crystal prototype sensor, the cleaned IDE is placed inside the reaction vial of ZIF-70 MOF. The MOF synthesis reaction parameters were kept the same as described in §1.1. After the completion of the reaction, the IDE electrode was gently washed in the methanol solvent to avoid any stress-related cracking and then dried at room temperature. The IDE electrode was weighed before and after the crystal growth step to keep track of the amount of deposited ZIF-70 amount, which was later used to quantify the adsorbed iodine amount.

**Inkjet printing:** The MOF 'ink', comprising an isopropyl alcohol (IPA) suspension of fine MOF particles without any additives, were printed using a commercially available printing system (Jetlab-II, MicroFab Technologies Inc.) which can deposit droplets within ±5 µm accuracy. To print MOF droplets without any satellite droplet, piezoelectric nozzle with an 80-µm orifice diameter was employed. The dispenser was plugged with a pneumatic pressure control system to control the back pressure so as to reduce the isopropanol solvent evaporation at the tip of the nozzle (the fast evaporation at the tip could lead to nozzle blockage). Figure S1 shows the dynamic shadowgraph images of the smooth MOF droplet formation process. The in-flight diameter of the falling droplet is ~50 µm. The IDE patterned substrate (Micrux interdigitated electrodes) was placed onto the bed of the printer. During printing, the substrate and printhead were maintained at room temperature of 21 °C. A rectangular array of droplets, according to the area of the IDE electrodes, were printed at a droplet spacing of 20 µm. Iodine sensing MOF based devices with a range of film thicknesses were produced by changing the print pass of the printhead.

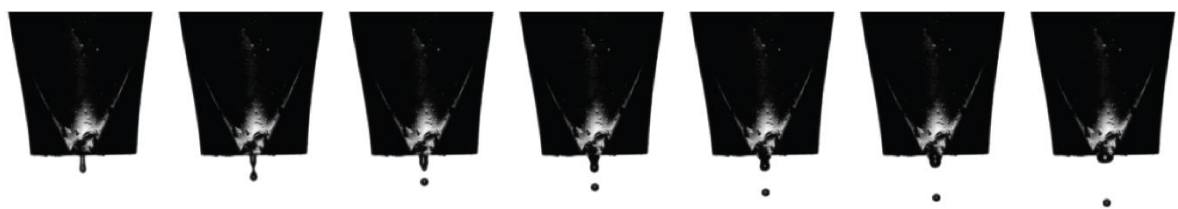

Figure S1: High-speed photography images of the printing process of an exemplar ZIF-70 MOF solution. The size of the nozzle orifice is 80 µm. The time interval between two adjacent frames is 48 µs.

**Iodine Exposure**

For iodine experiments, the MOF deposited IDE electrode and an excess of 50 mg of dry iodine was placed inside a 500 mL glass bottle. The sealed glass bottle was then placed inside an oven at 70 °C for 30 mins and later cooled down to room temperature. For each MOF, three IDEs were prepared to ensure the repeatability of data. MOF sample with the highest iodine sensitivity was chosen to further carry out cyclic adsorption and desorption measurement for 3 cycles to confirm its reversibility. To do so, the sample was heated at 70 °C overnight.

## 1.3 Electrical response

The electrical response from the IDE sensor was recorded at room temperature at 35% RH (relative humidity) using the HIOKI IM3536 LCR meter in the frequency range of 4 Hz to 8 MHz at 1 V. To measure the sensitivity of the sample IDE at any point in time, the impedance, capacitance and phase angle parameters were obtained in parallel as a function of frequency. For in-situ experiments, a 467.6 ppb and 8.35, 16.7, 25.05, 33.40 ppm level $I_2$ environment was maintained by placing meshed iodine inside an in-house closed chamber alongside the sample IDE to simulate the real-time application. Subsequently, the impedance measurements were continuously collected at 10 Hz (1 V) as well as over the frequency range of 4 Hz to 8 MHz.

## 1.4 X-ray diffraction (XRD)

The powder XRD pattern for the different MOF samples was determined using the Rigaku MiniFlex benchtop X-ray diffractometer at a scan rate of 0.2°/min with a step size of 0.05°. Before the data collection, the MOF samples were pre-evacuated in a vacuum chamber at 100 °C overnight to minimize solvent effect.

## 1.5 Fourier-transform infrared (FTIR) spectroscopy

The FTIR spectra for the MOF samples were recorded using the Nicolet-iS10 FTIR spectrometer equipped with an attenuated total reflectance (ATR) sample accessory in the mid-IR region (650–4000 $cm^{-1}$) at a spectral resolution of 0.5 $cm^{-1}$ after collecting the background spectrum using the identical parameters. The impact of iodine adsorption and desorption on infrared vibrational modes of MOF at room temperature as well as with time-dependent heating effect was studied by ATR-FTIR.

## 1.6 Raman spectroscopy

The Raman spectra were collected using the Bruker MultiRAM Raman spectrometer with sample compartment D418, equipped with a Nd-YAG-Laser (1064 nm) and a LN-Ge diode as a detector. The laser power used for sample excitation was 50 mW, and 64 scans were accumulated at a resolution of 1 cm$^{-1}$.

## 1.7 Thermogravimetric analyses (TGA)

The thermal stability of the MOF specimens was measured using the TGA-Q50 (TA Instruments) equipped with an induction heater (max temperature 1000 °C) and platinum sample holder under an $N_2$ inert atmosphere. The samples were heated at a rate of 10 °C/min from 30 to 800 °C.

## 1.8 UV−Visible diffuse reflectance spectroscopy (DRS)

The absorption spectra for samples were obtained using the 2600 UV−Vis spectrophotometer (Shimadzu) in the wavelength range of 200-1400 nm, equipped with an integrating sphere. The diffused reflectance spectra (DRS) were measured and converted using the Kubelka−Munk (KM) transformation to estimate the optical band gaps.

## 1.9 Optical microscopy and surface profilometry

Alicona profilometer was used to measure the surface texture such as the thickness of the deposited MOF layer. The surface topography was characterized by the infinite focus microscopy technique (Alicona Infinite Focus 3D profilometer) using the 5× optics on the profilometer.

## 2. Powder X-ray diffraction (PXRD) of different MOFs

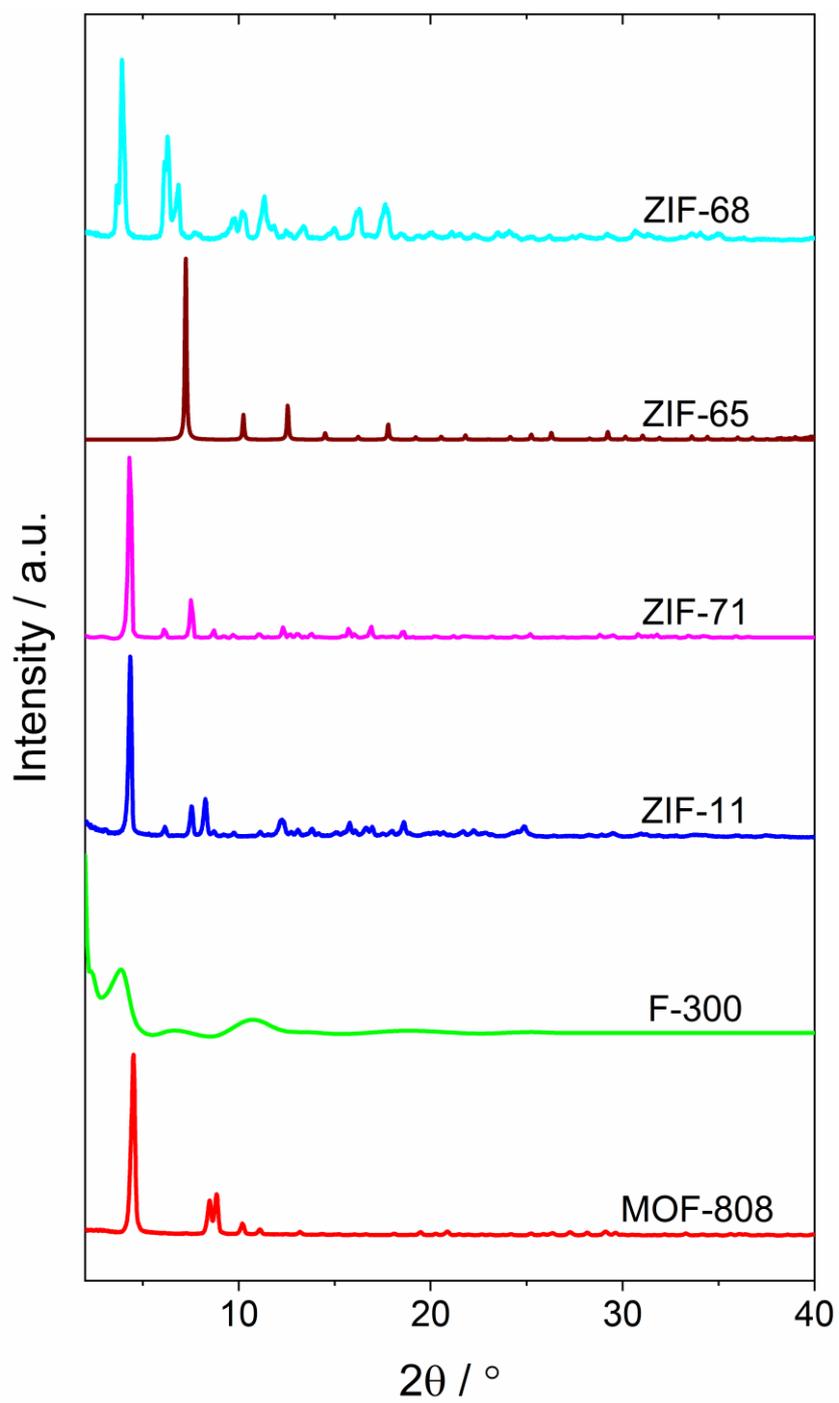

Figure S2: Normalized PXRD patterns of activated MOF powder samples.

## 3. IDE sensors integrating different MOF samples

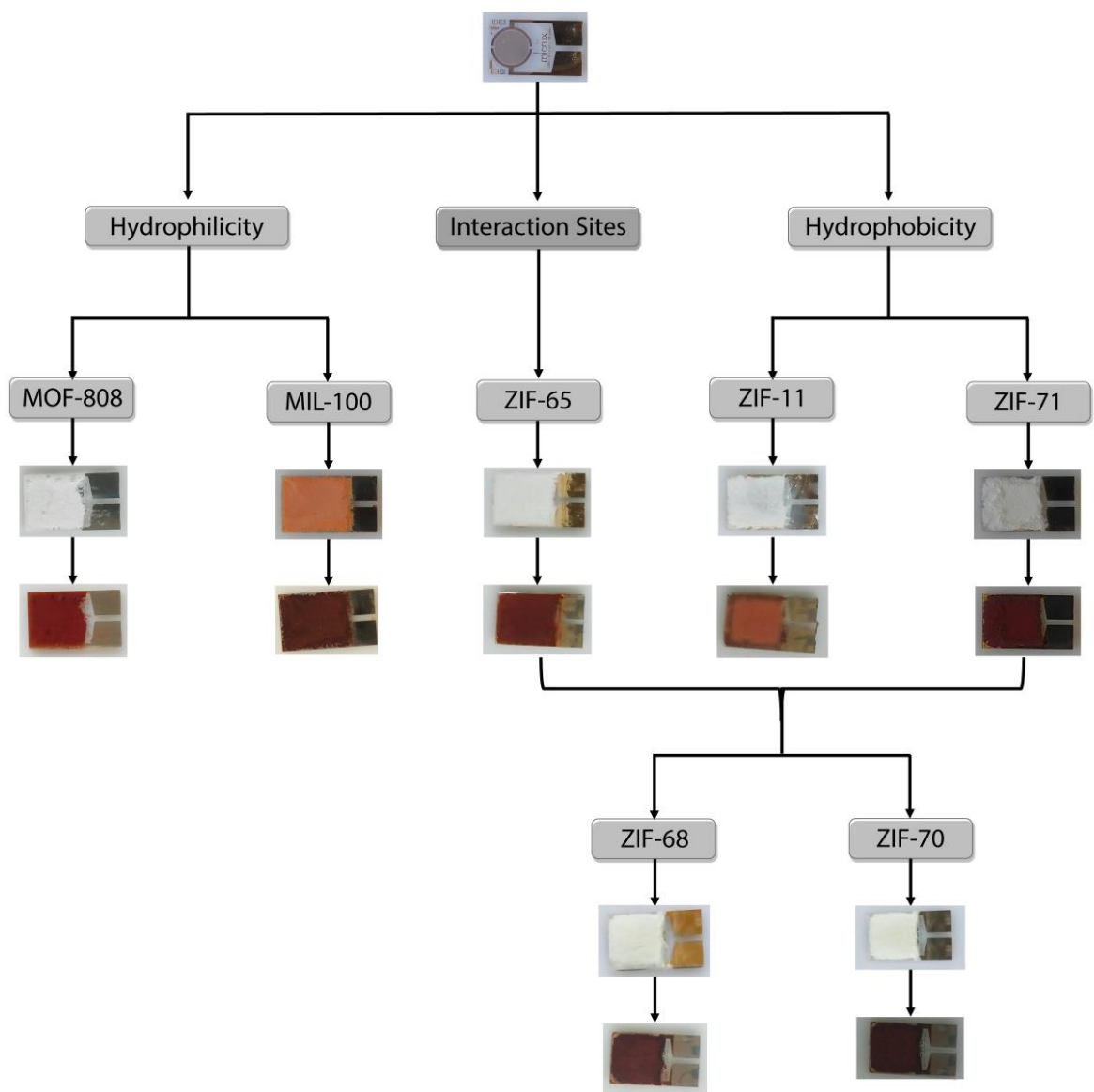

Figure S3: Photographs of prototype MOF@IDE sensors, before and after iodine exposure tests (top vs. bottom images of each MOF sample).

Table S1: Iodine adsorption dependent percentage change in the MOF weight.

| MOF | Mass Change / % |
|---|---|
| MOF-808 | 50.0±12.0 |
| F300 | 48.2±2.0 |
| ZIF-11 | 10.7±4.8 |
| ZIF-71 | 26.6±6.0 |
| ZIF-65 | 47.0±1.2 |
| ZIF-68 | 26.8±7.0 |
| ZIF-70 | 91.0±8.2 |

## 4. MOF@IDE sensor response

## 4.1 Hydrophilicity effect of MOF-808 and F300

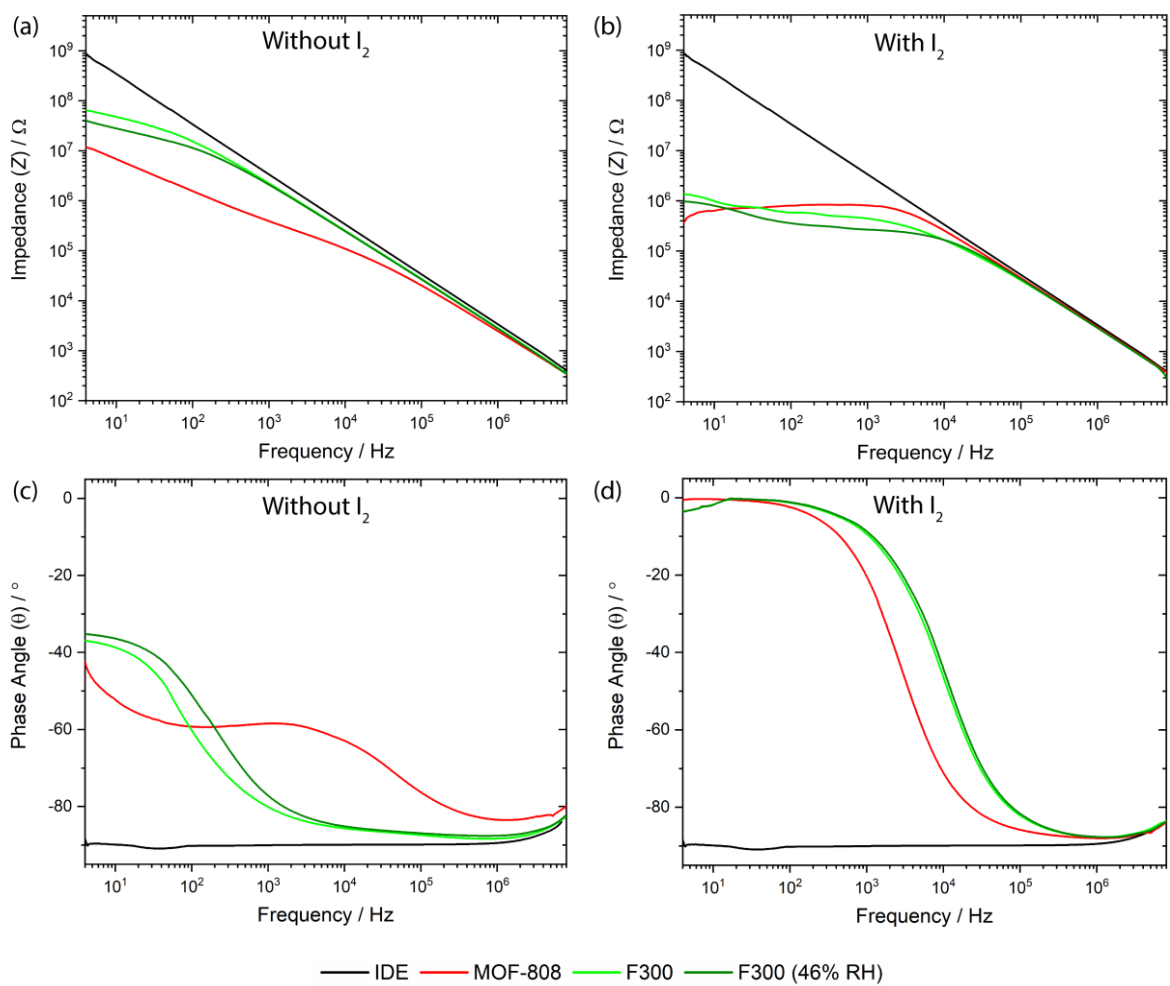

Figure S4: Effect of hydrophilicity on sensor impedance and phase angle: (a), (c) before and (b), (d) after iodine adsorption, respectively.

## 4.2 Hydrophobicity effect of ZIF-11 and ZIF-71

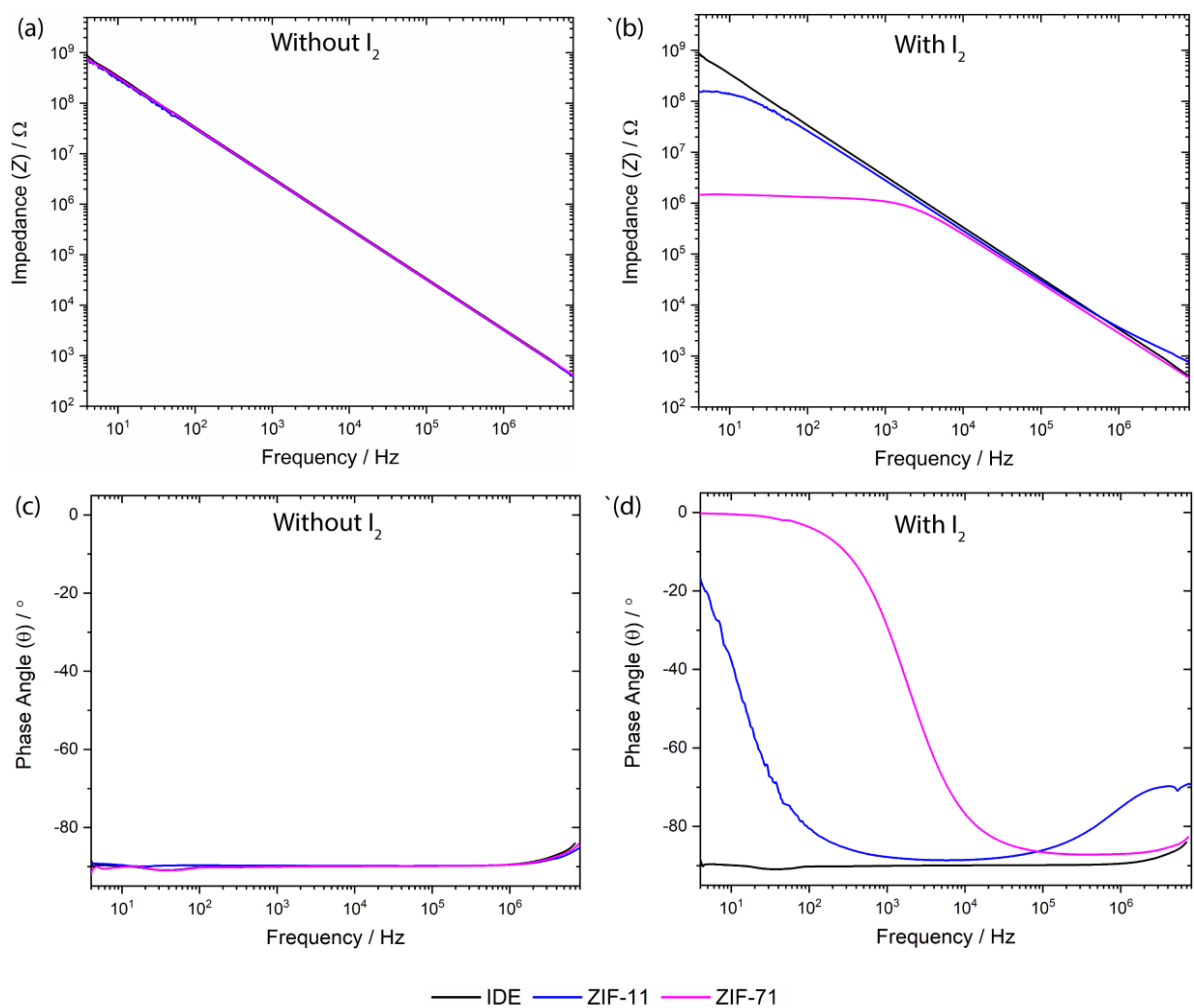

Figure S5: Hydrophobicity dependent alterations in sensor impedance and phase angle: (a), (c) before and (b), (d) after iodine adsorption, respectively.

## 4.3 Interaction sites effect of ZIF-65

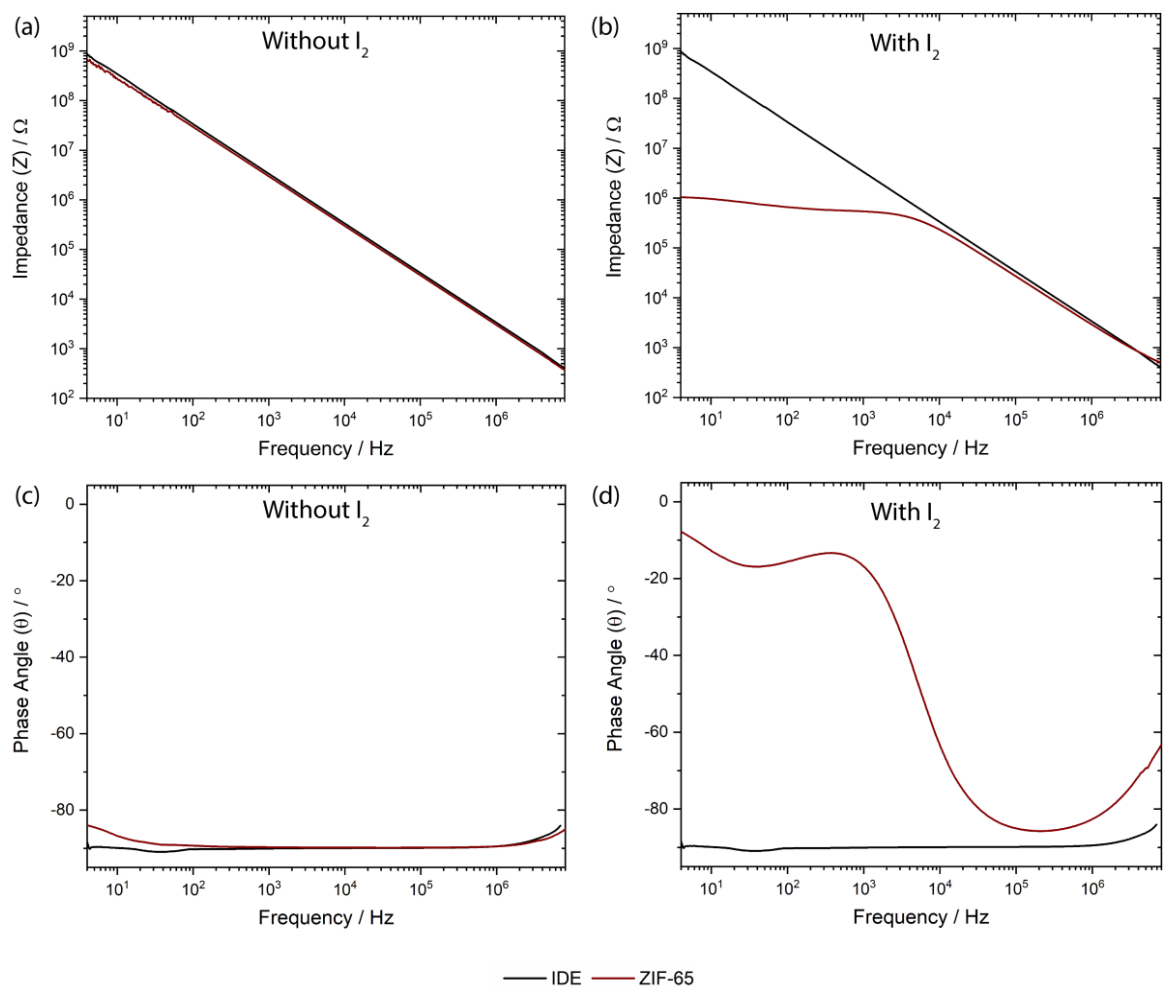

Figure S6: Interaction site dependent shift in sensor impedance and phase angle: (a), (c) before and (b), (d) after iodine adsorption, respectively.

## 4.4 Interaction sites with optimal hydrophobicity effects of ZIF-65 and ZIF-70

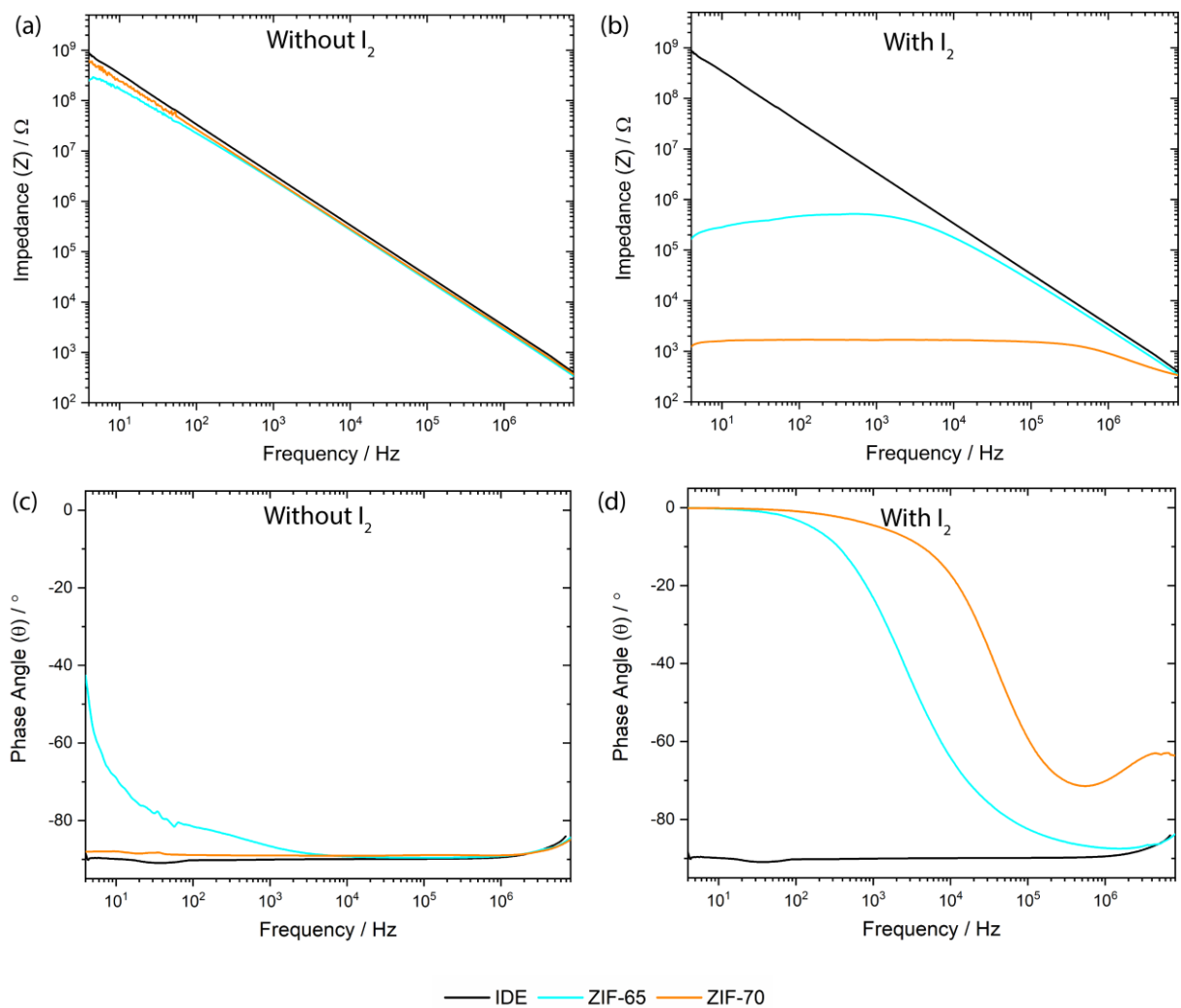

Figure S7: Combined effect of interaction site and hydrophilicity-hydrophobicity on sensor impedance and phase angle: (a), (c) before and (b), (d) after iodine adsorption, respectively.

## 5. Change in MOF@IDE response

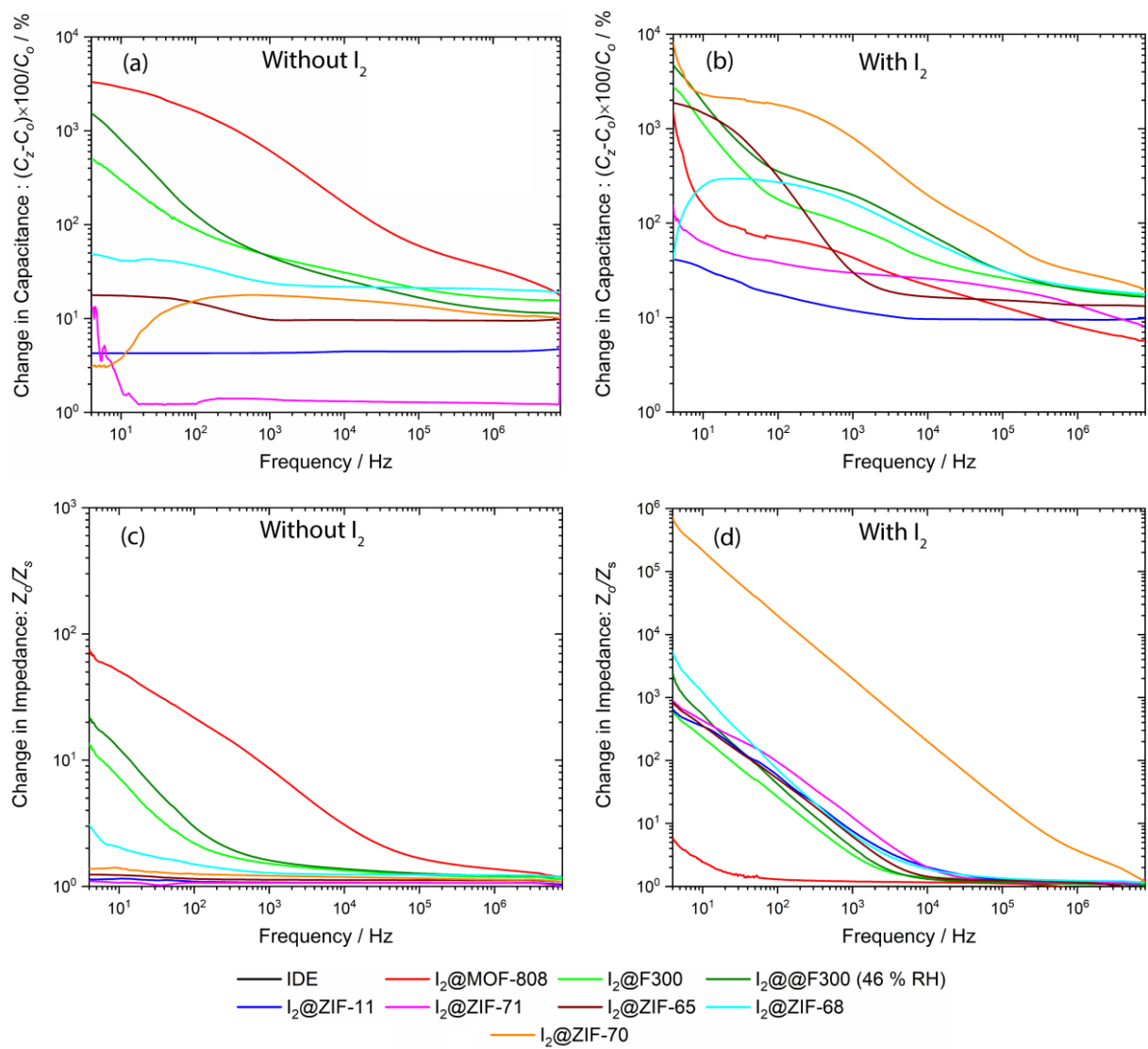

Figure S8: Variations in the electrical response of different MOF@IDEs: before and after iodine adsorption. (a), (b) Changes in capacitance and (c), (d) in impedance, as a function of frequency.

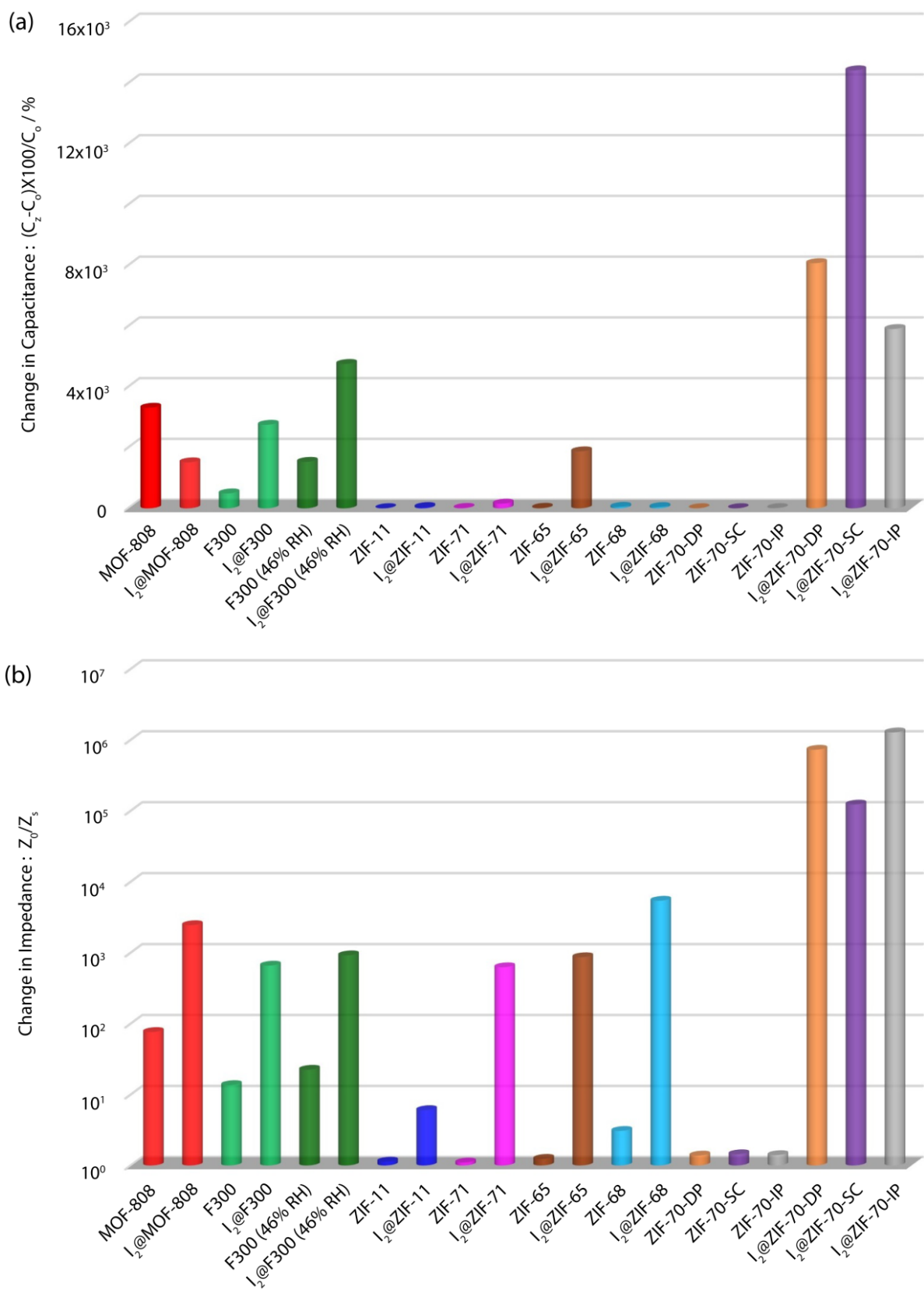

Figure S9: Comparative plots of the change in MOF@IDE output parameters determined at 4 Hz: (a) Percentage change in capacitance, and (b) ratio of sample impedance relative to an 'empty' IDE (no MOF). Note: DP = drop casting, SC = single-crystal and IP = inkjet printing.

Table S2: Enhancement in capacitance and impedance ratio of different drop-casted MOF@IDEs, before and after the I$_2$ exposure.

| MOF Name | n-Fold in Enhancement | |
|---|---|---|
| | **Capacitance** $(C_z-C_o) \times 100/C_o$ | **Impedance** $Z_o/Z_s$ |
| **MOF-808** | 3,311.1 | 76.1 |
| **I$_2$@MOF-808** | 1,512.8 | 2,439.2 |
| **F300** | 489.6 | 13.6 |
| **I$_2$@F300** | 2,751.2 | 656.3 |
| **F300 (46% RH)** | 1,530.5 | 22.4 |
| **I$_2$@F300 (46% RH)** | 4,754.9 | 916.8 |
| **ZIF-11** | 4.3 | 1.1 |
| **I$_2$@ZIF-11** | 41.4 | 6.0 |
| **ZIF-71** | 13.0 | 1.1 |
| **I$_2$@ZIF-71** | 153.5 | 622.3 |
| **ZIF-65** | 17.8 | 1.2 |
| **I$_2$@ZIF-65** | 1,876.3 | 862.2 |
| **ZIF-68** | 48.9 | 3.1 |
| **I$_2$@ZIF-68** | 41.3 | 5,394.3 |
| **ZIF-70** | 3.0 | 1.4 |
| **ZIF-70-SC** | 4.2 | 1.4 |
| **ZIF-70-IP** | 6.6 | 1.4 |
| **I$_2$@ZIF-70** | 8,074.6 | 725,042.7 |
| **I$_2$@ZIF-70-SC** | 14,422.2 | 122,347.9 |
| **I$_2$@ZIF-70-IP** | 5,904.5 | 1,275,056.9 |

## 6. Guest-dependent MOF@IDE response

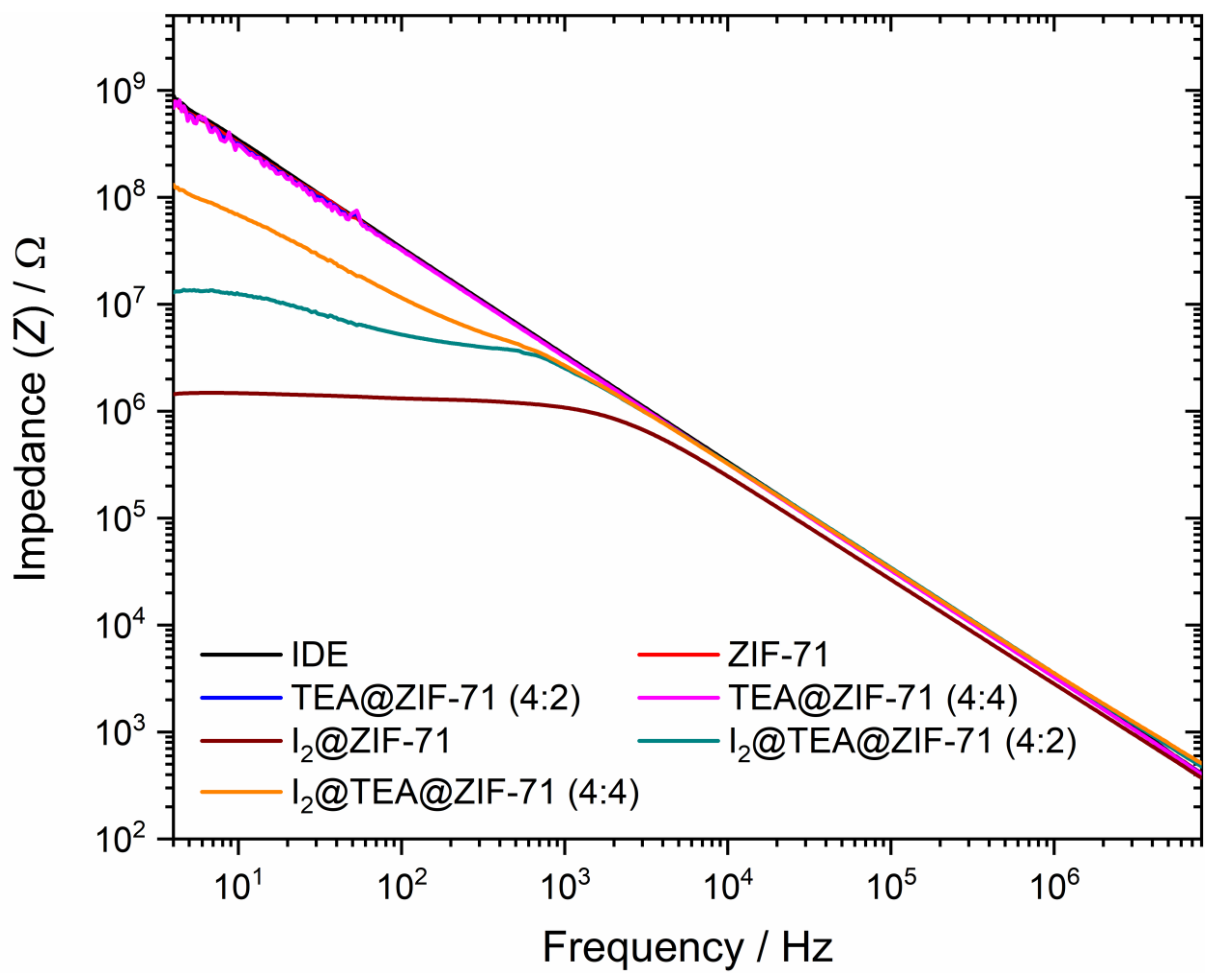

Figure S10: Effect of different amount of guest encapsulation in ZIF-71 framework on iodine sensing. The ratio of linker to TEA was varied during the ZIF-71 synthesis and expressed in label as ZIF-71(L: G), where L and G are the molar concentrations of the MOF linker and triethylamine (TEA) guest, respectively.

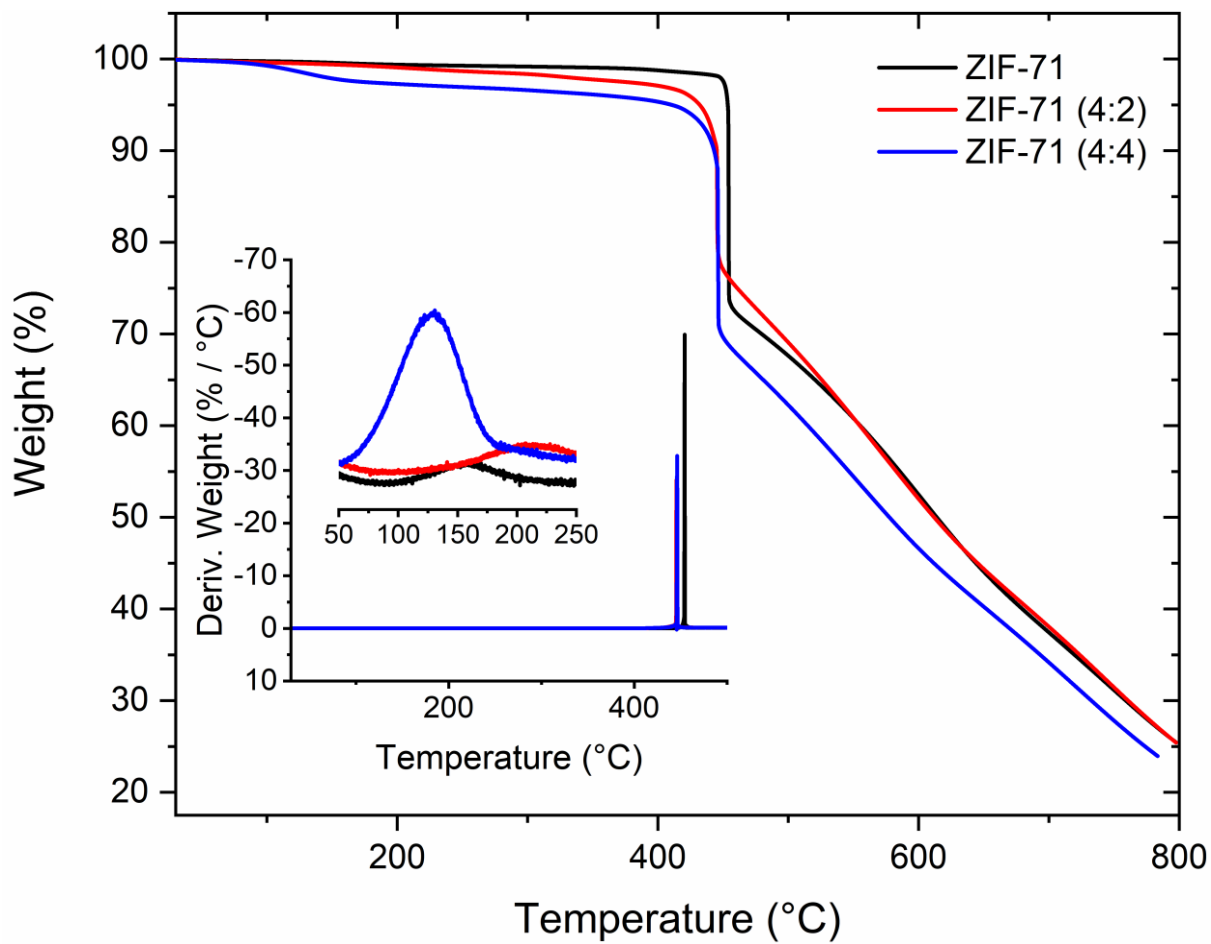

Figure S11: TGA analysis to show the presence of triethylamine (TEA) guest in ZIF-71 pores. The ratio of linker to TEA was varied during the ZIF-71 synthesis and expressed in label as ZIF-71(L: G), where L is linker and G is TEA Guest.

## 7. MOF@IDE sensitivity in the presence of different solvent vapors

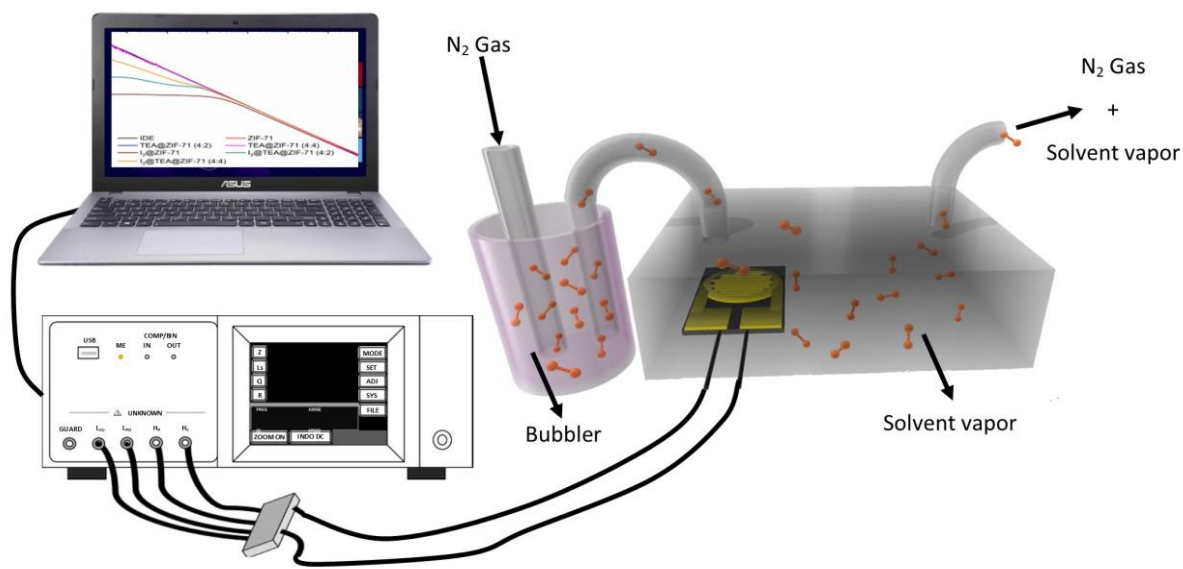

Figure S12: Customized setup designed for in-situ testing of various saturated vapour effects on the electrical impedance (*via* LCR meter) of the ZIF-70 MOF.

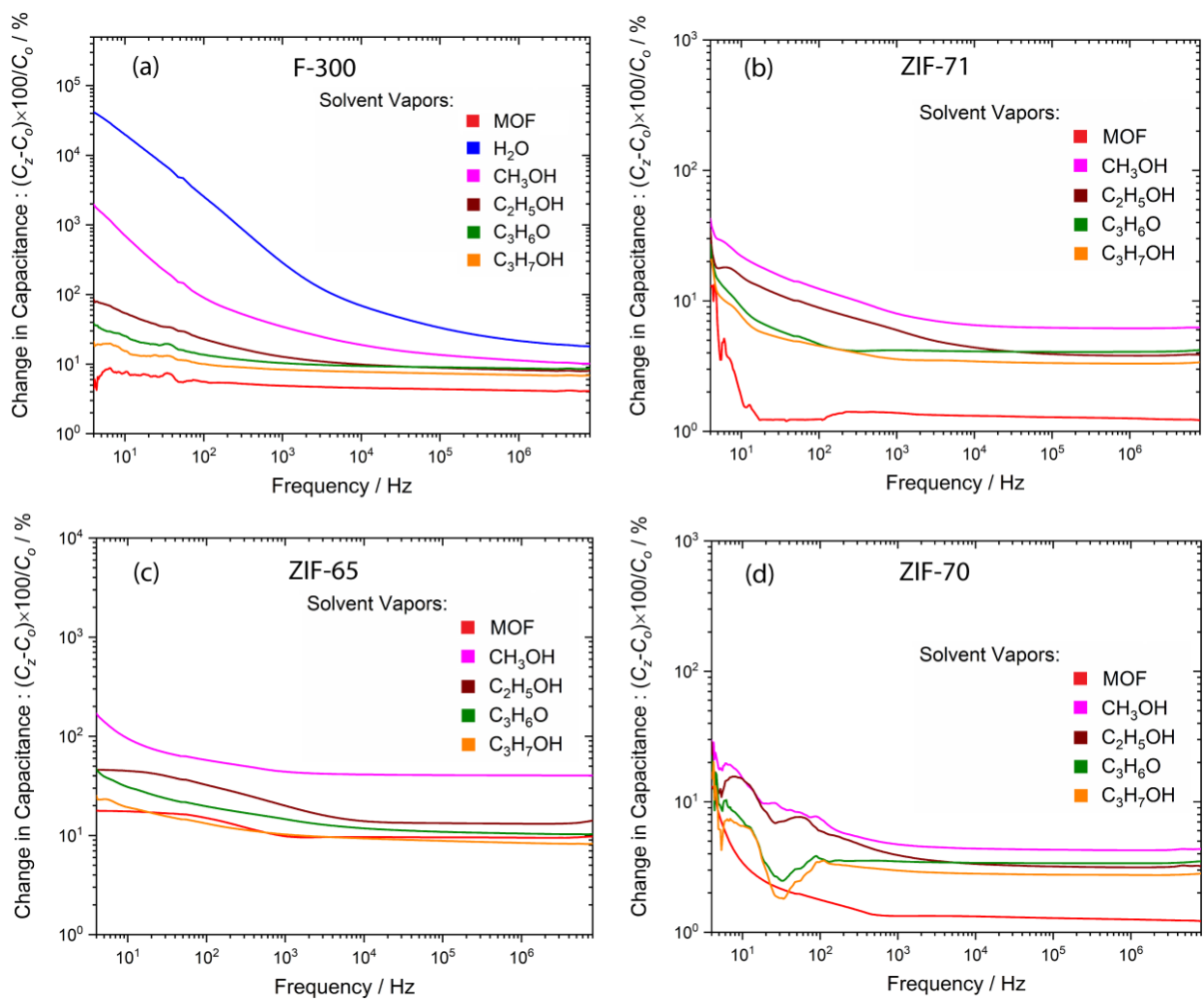

Figure S13: Solvent vapor dependent change in capacitance response for different MOF prototype sensors: (a) F300, (b) ZIF-71, (c) ZIF-65 and ZIF-70 MOFs.

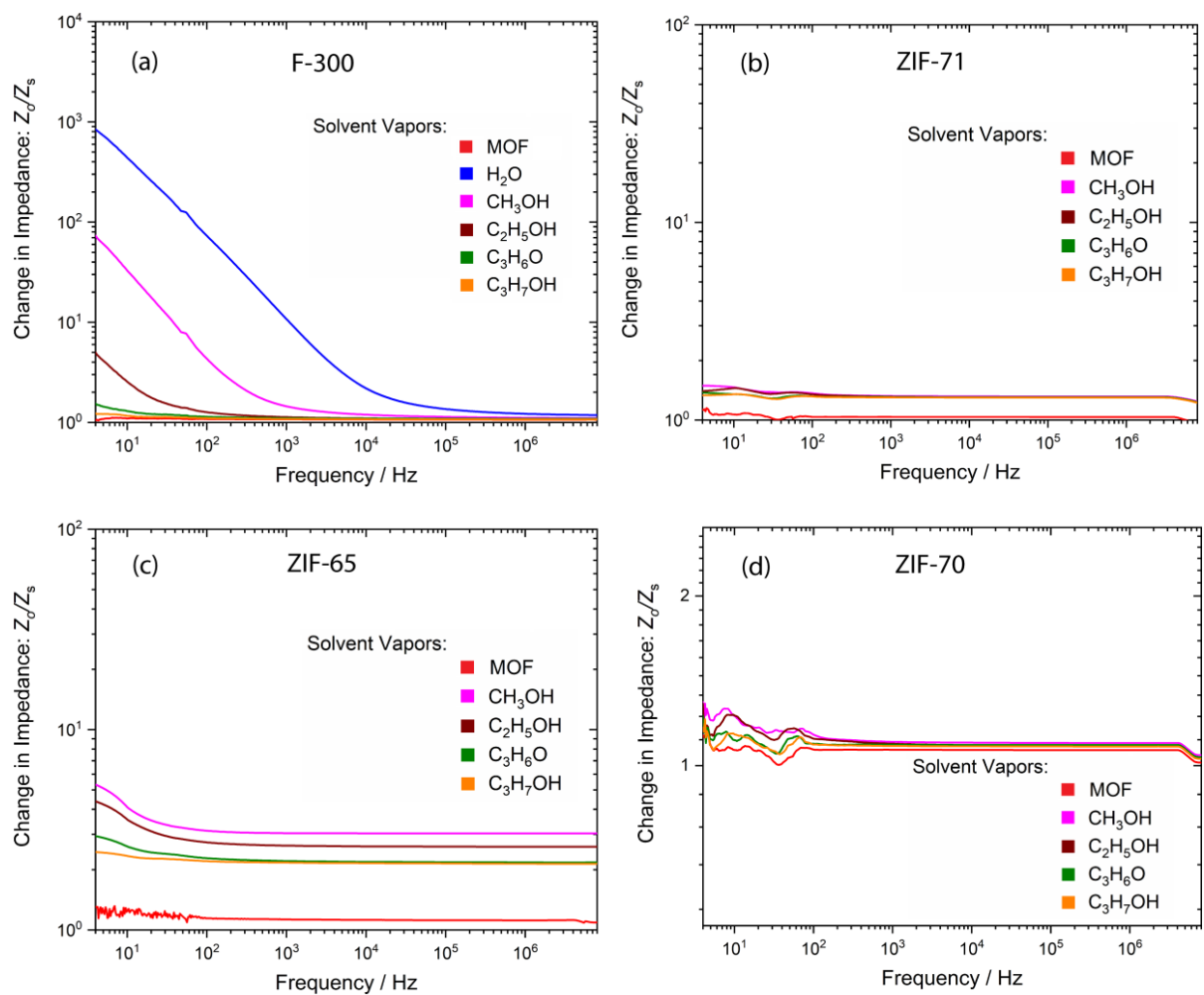

Figure S14: Solvent vapor dependent change in impedance response for different MOF prototype sensors: (a) F300, (b) ZIF-71, (c) ZIF-65 and ZIF-70 MOFs.

Table S3: A comparison between $I_2$ and various solvent-dependent changes in the impedance ratio of drop-casted ZIF-70 MOF.

| Target Molecules@ZIF-70 | Change in Impedance ($Z_o/Z_s$) |
|---|---|
| $I_2$@ZIF-70 | 725,042.7 |
| Methanol | 1.3 |
| Ethanol | 1.3 |
| Acetone | 1.2 |
| Isopropanol | 1.2 |

## 8. Adsorption-desorption response of ZIF-70 in the presence of solvent vapors

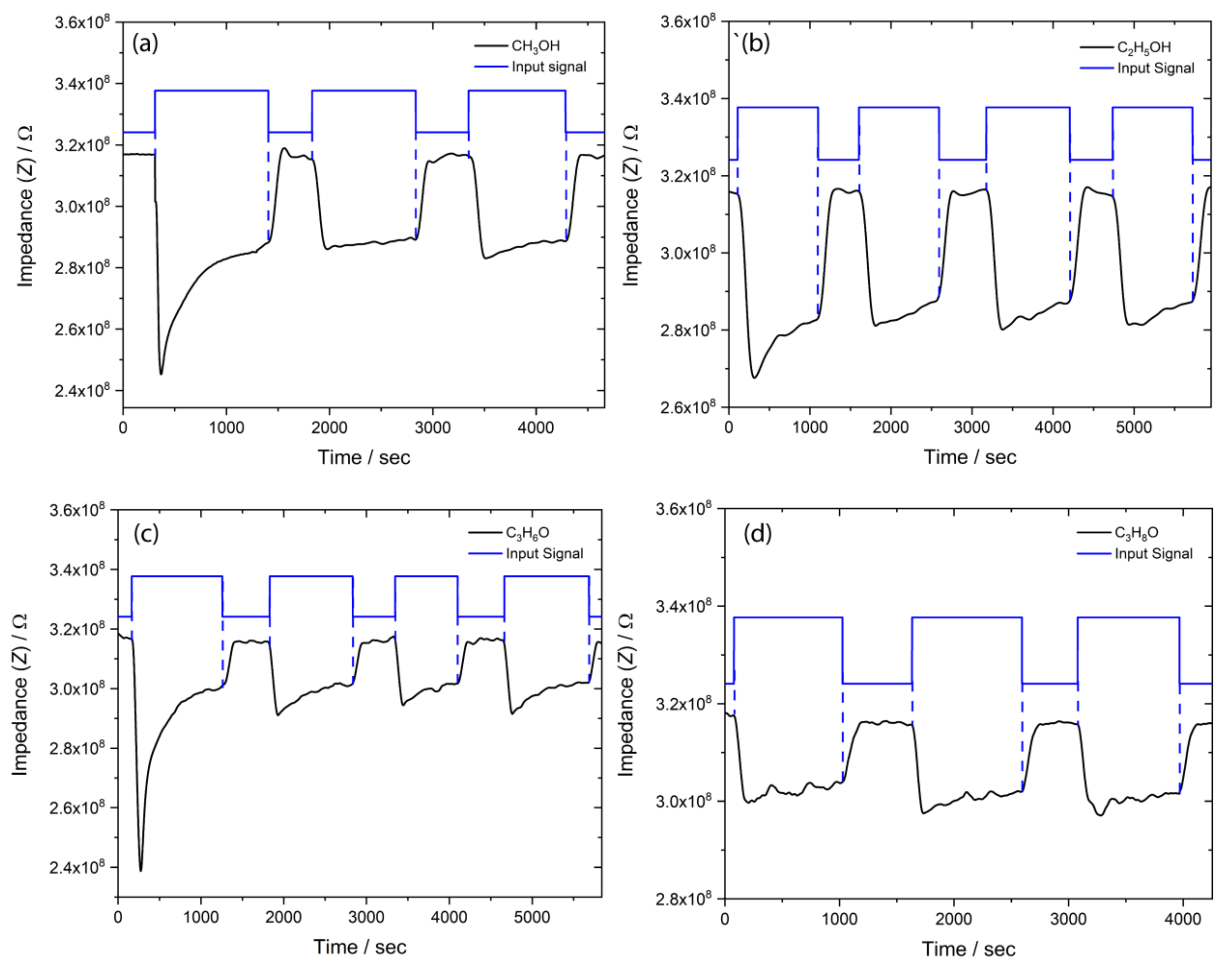

Figure S15: Solvent adsorption-desorption dependent cyclic impedance response for ZIF-70 MOF at 10 Hz, when exposed to (a) methanol, (b) ethanol, (c) acetone, and (d) isopropanol.

## 9. Single-crystal ZIF-70 MOF@IDE sensitivity in the presence of iodine

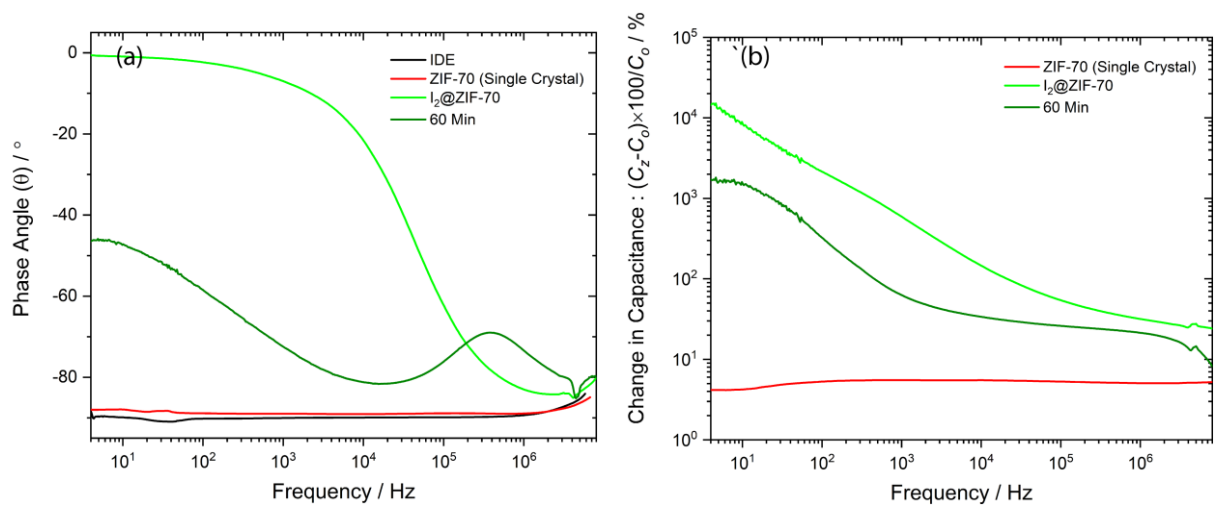

Figure S16: Variations of (a) phase angle and (b) capacitance in single-crystal ZIF-70 prototype sensor, before and after the iodine exposure.

## 10. Inkjet-printed ZIF-70 sample layer thickness by optical profilometry

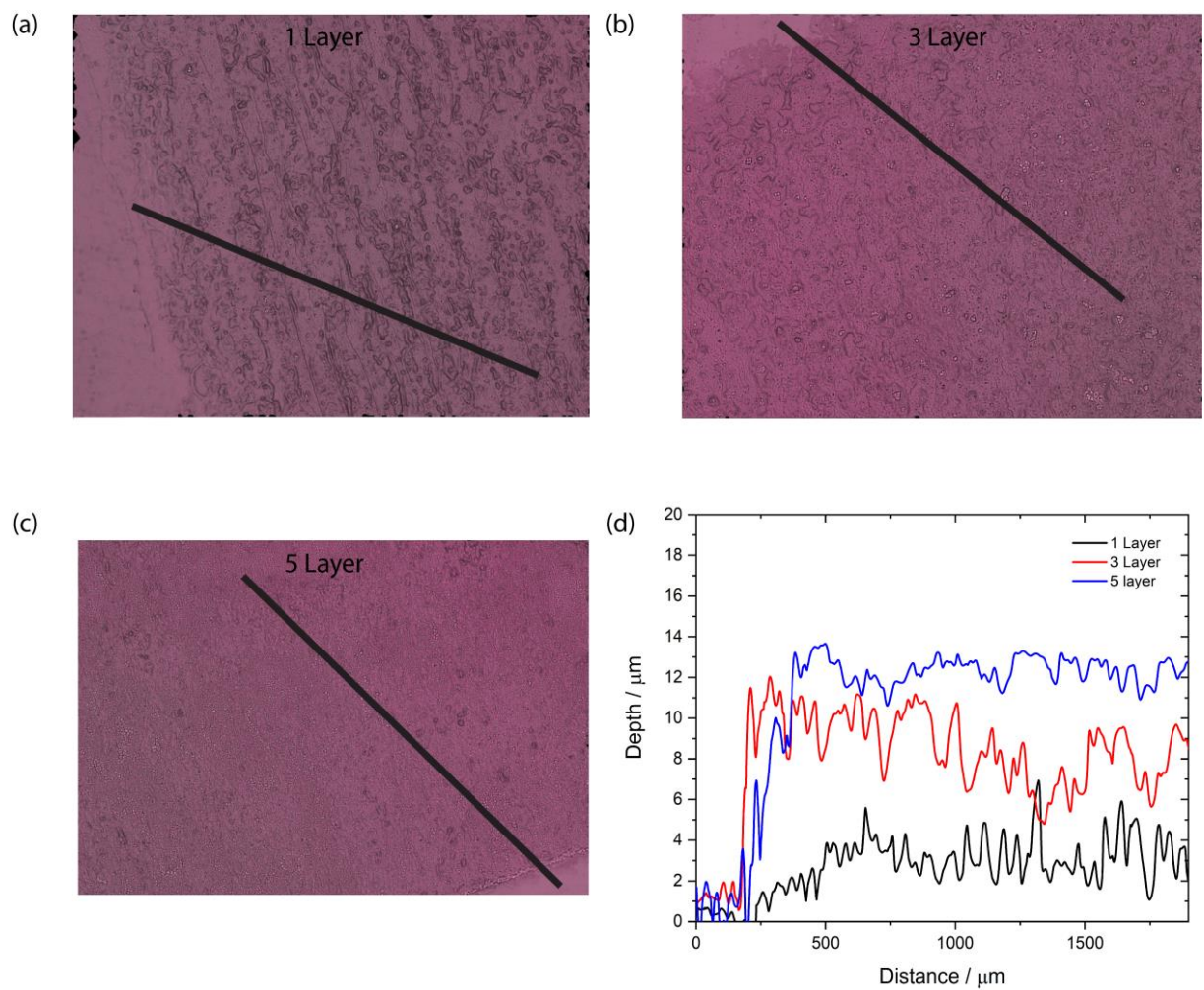

Figure S17: Optical images of inkjet-printed ZIF-70 layers: (a) 1 layer, (b) 3 layers and (c) 5 layers. (d) Plots of ZIF-70-layer height with respect to the distance associated with the black line marked in the corresponding optical images.

## 11. Inkjet-printed ZIF-70 MOF@IDE sensitivity in ppm and ppb levels

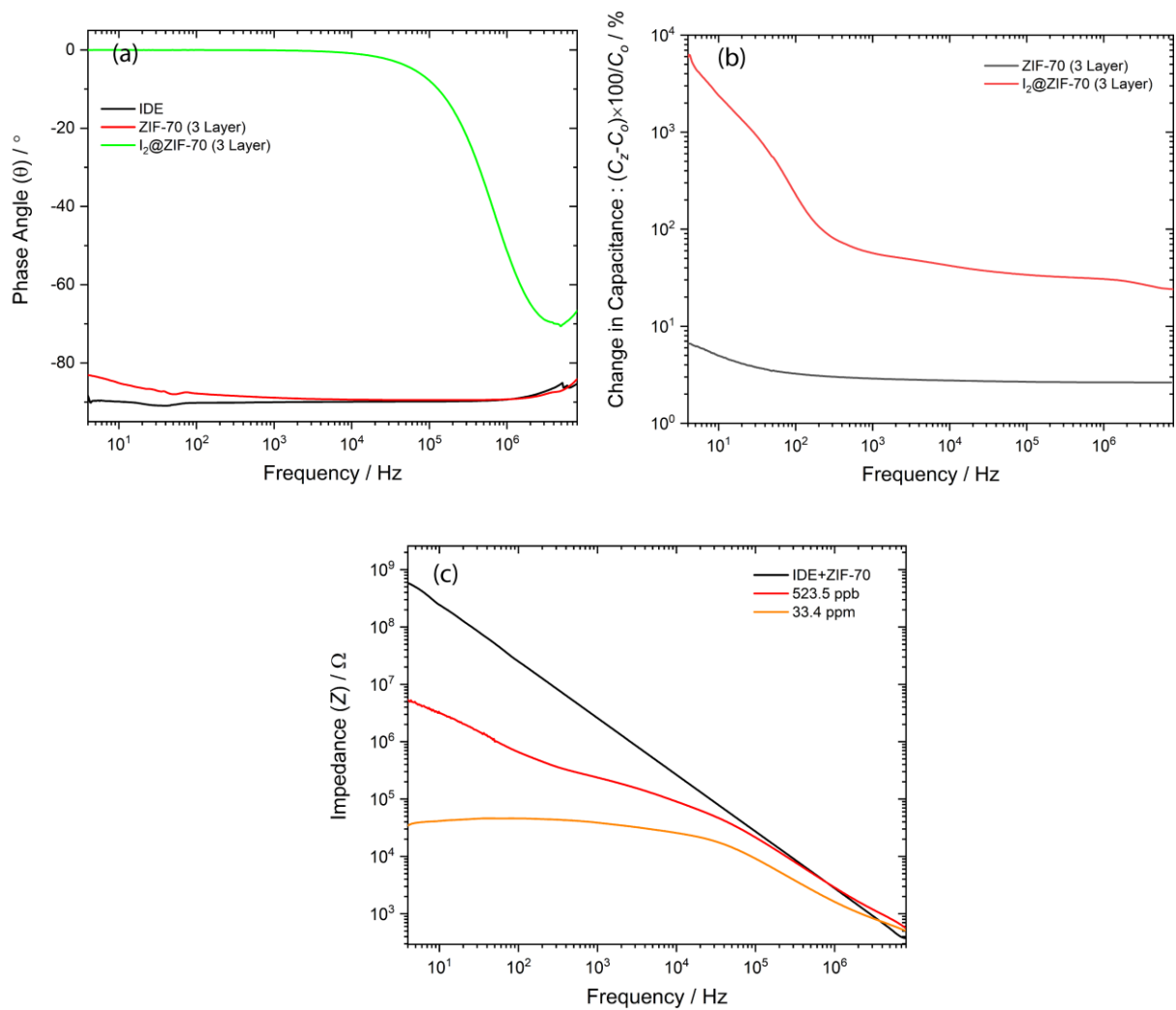

Figure S18: (a) Phase angle and (b) change in capacitance of 3-layer-ZIF-70@IDE, before and after iodine exposure. (c) Impedance sensitivity of 3-layer-ZIF-70@IDE prototype sensor performance at ppm and ppb concentration levels.

## 12. Characterization of ZIF-70 before and after I$_2$ uptake

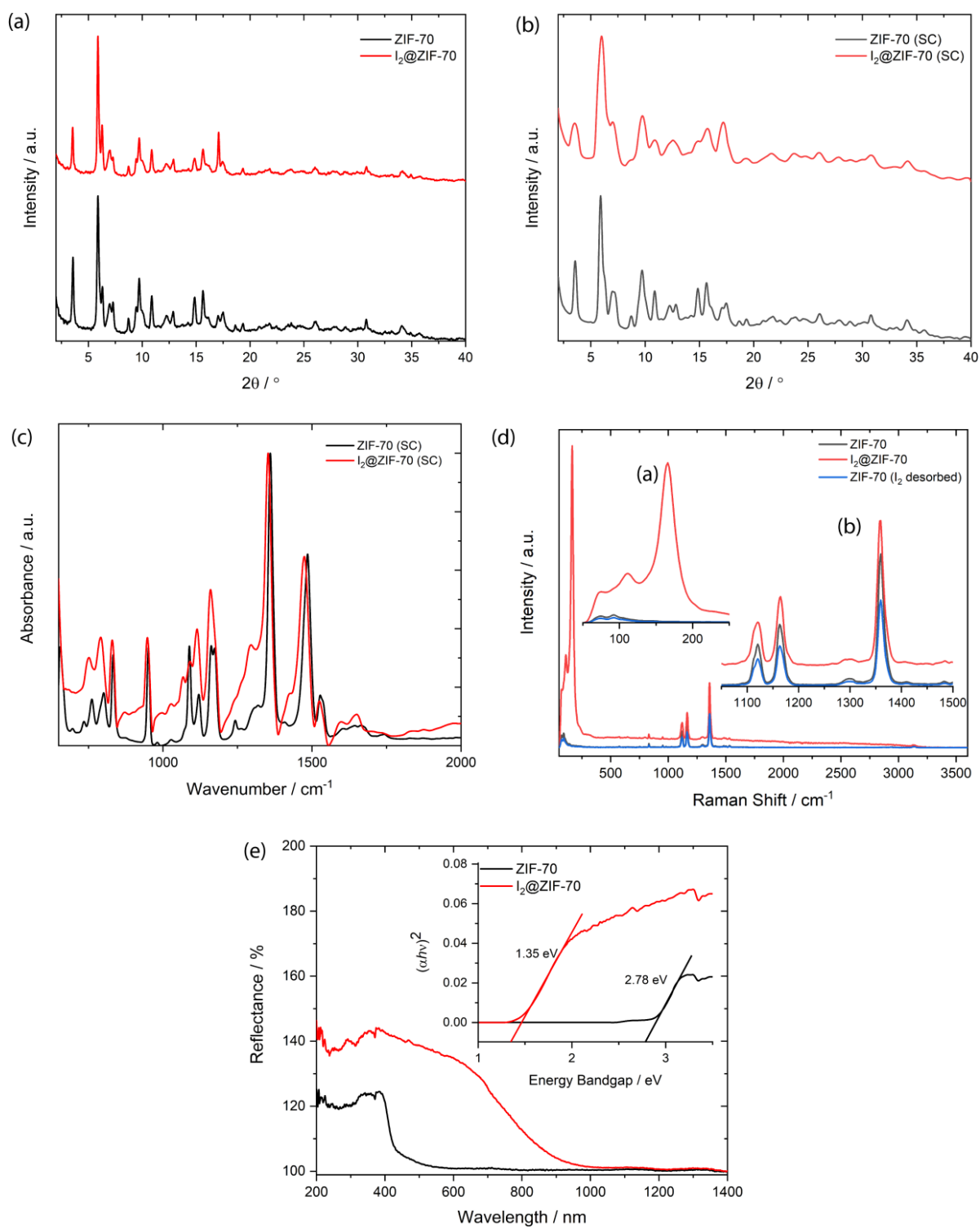

Figure S19: Material characterization of ZIF-70 MOF, before and after iodine exposure: XRD patterns of (a) powder and (b) single crystal of ZIF-70. (c) FTIR spectrum of ZIF-70 single crystal, and (d) Raman spectra of ZIF-70 powder. (e) UV-Vis diffuse reflectance spectra (DRS) with an inset of band-gap determination *via* Kubelka-Munk (KM) method.

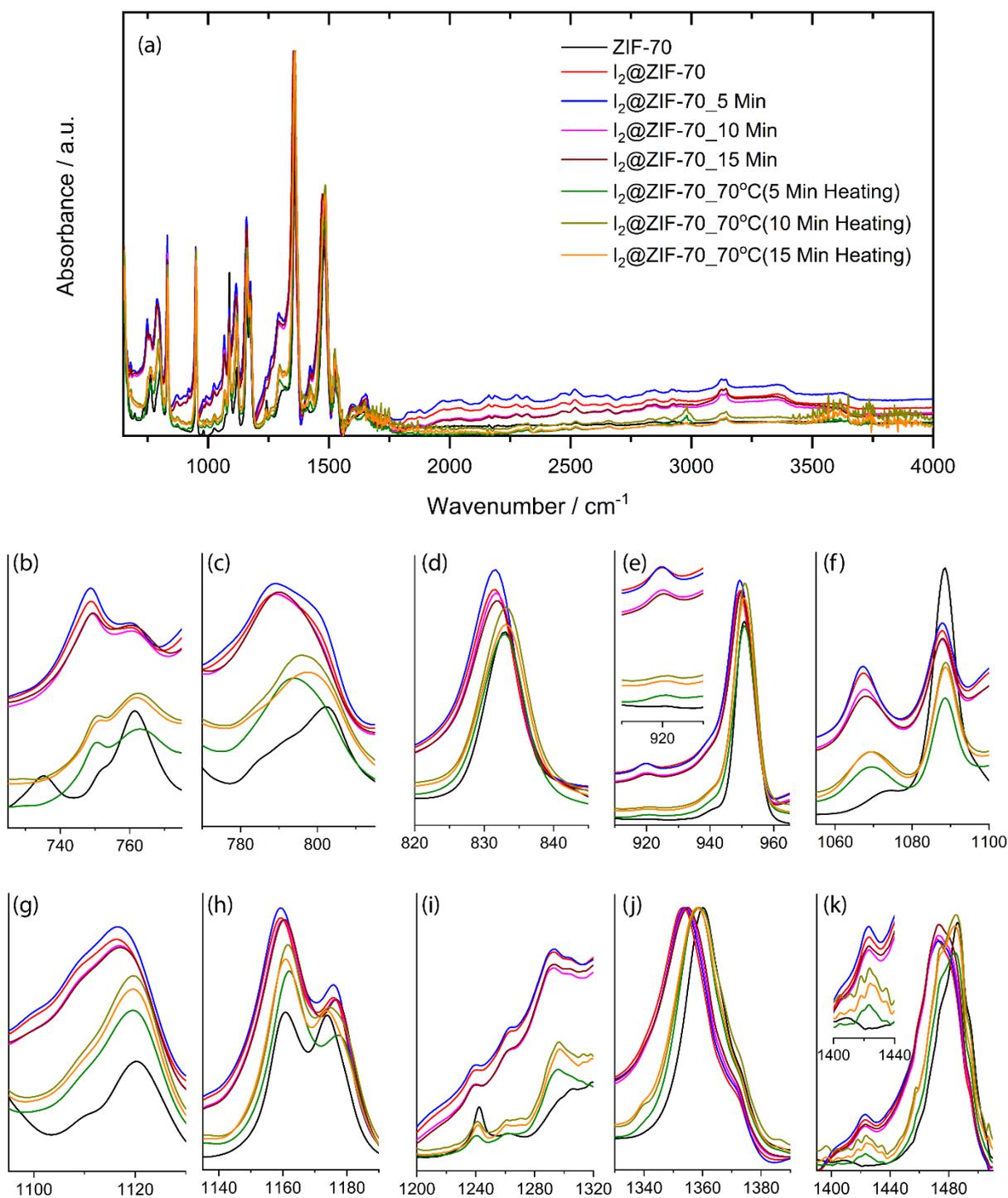

Figure S20: Time- and temperature-dependent iodine desorption analysis for ZIF-70 MOF using the FTIR spectra.

## 13. Sensor performance data

Table S4: Enhancement in the impedance ratio of the ZIF-70@sensor prepared using drop-casting, single crystal, and inkjet printing methods. The impedance measurements were performed under the DC and AC frequencies.

| Method | DC | AC |
| --- | --- | --- |
| Drop-casted | $9.1 \times 10^8$ | $7.2 \times 10^5$ |
| Single crystal | $5.9 \times 10^8$ | $1.2 \times 10^5$ |
| Inkjet printing | $2.8 \times 10^9$ | $1.3 \times 10^6$ |

Table S5: Comparison of the sensor sensitivity between this work and other related investigations on IDE-based iodine sensors. Literature values were extracted from the published studies cited below.

| Materials | Direct Current (DC) | Alternative Current (AC) | Reference |
|---|---|---|---|
| **ZIF-70** | $2.84 \times 10^9$ | 54,973 | This work |
| *MFM-300 (Al)* | < 1,000,000 | < 10 | *ACS Appl. Mater. Interfaces* **11**, 27982-27988 (2019). |
| *ZIF-8* | < 100,000 | < 100 | *ACS Appl. Mater. Interfaces* **9**, 44649-44655 (2017). |
| *Polyacetylenic film* | < 1,000 | - | Sens. Actuators B **129**, 171 (2008). |
| *Ag-mordenite* | < 300 | < 5 | *Microporous Mesoporous Mater.* **280**, 82−87 (2019) |